\documentclass[twocolumn,floatfix,superscriptaddress,amsmath,showpacs,showkeys,aps,pre,eqsecnum]{revtex4-1}
\usepackage[utf8]{inputenc}
\usepackage[final]{graphicx}
\usepackage{bm}
\usepackage[normalem]{ulem}
\usepackage[usenames]{color}
\usepackage{times}
\usepackage{verbatim}
\usepackage{xcolor}
\usepackage{hyperref}
\usepackage{wrapfig}
\usepackage{subfigure}
\DeclareGraphicsExtensions{.eps}

\begin{document}
\title{Conditional probabilities in  multiplicative noise processes}
\author{Miguel V. Moreno}
\affiliation{Departamento de F{\'\i}sica Te\'orica,
Universidade do Estado do Rio de Janeiro, Rua S\~ao Francisco Xavier 524, 20550-013,  Rio de Janeiro, RJ, Brazil.}
\author{Daniel G. Barci}
\affiliation{Departamento de F{\'\i}sica Te\'orica,
Universidade do Estado do Rio de Janeiro, Rua S\~ao Francisco Xavier 524, 20550-013,  Rio de Janeiro, RJ, Brazil.}
\author{Zochil Gonz\'alez Arenas}
\affiliation{Departamento de Matem\'atica Aplicada, IME, Universidade do Estado do Rio de Janeiro, Rua S\~ao Francisco Xavier 524, 20550-013,  Rio de Janeiro, RJ, Brazil}
\date{\today}

\begin{abstract}
We address the calculation of transition probabilities in multiplicative noise stochastic differential equations using a path integral approach. 
We show the equivalence between the conditional probability and the propagator of a quantum particle with  variable mass. Introducing a {\em  time reparametrization}, we are able to transform the problem of multiplicative noise fluctuations 
into an equivalent additive one.  We illustrate the method by showing the explicit analytic computation of the conditional probability 
of a harmonic oscillator in a nonlinear multiplicative environment. 
\end{abstract}

\maketitle

\section{Introduction}
Stochastic dynamics driven by multiplicative noise is, by now, a common arena to model complex behavior~\cite{gardiner,vanKampen}. 
 In fact, several very different areas of knowledge have benefited from stochastic modeling, such as physics and 
 chemistry~\cite{Poschel}, biology and ecology~\cite{Murray} and even  economy and social sciences~\cite{Mantegna, Bouchaud}.
From a physical perspective, multiplicative stochastic dynamics is one of the possible approaches to enhance our understanding of out-of-equilibrium 
statistical mechanics~\cite{crooks1999,ZwanzigBook2001,seifert2008}.

Some typical examples of multiplicative noise processes are the diffusion of particles near a 
wall~\cite{Lancon2001,Lancon2002, Lubensky2007,Volpe2010,Volpe2011}, micromagnetic dynamics in ferromagnetic 
systems~\cite{GarciaPalacios1998, Aron2014, Arenas2018} and non-equilibrium transitions into absorbing states~\cite{Hinrichsen2000}. 
Moreover, among the very interesting stochastic phenomena described by multiplicative noise, there are two particular ones: noise-induced phase 
transitions~\cite{Parrondo1994,CastroWio1995,Sancho2003,Goldenfeld2015,BarciMiguelZochil2016} and stochastic 
resonance~\cite{Benzi1981,Parisi1983,Wio2007,Wio2002}, in which the interplay between nonlinearity and noise produces surprising  outputs. 

There are several theoretical approaches to deal with this type of processes. The choice of any particular one mainly depends on 
what kind of observables  we are interested in and what kind of calculation  technique we are willing to apply. For instance, 
the Langevin approach, consisting in a system of stochastic differential equations,  is the natural way to model a specific situation  
and to perform  numerical simulations~\cite{Sivak2014}. By the other hand, we can gain more insightful intuition by looking at the Fokker-Planck equation, 
which directly provides the time dependent probability distribution. Other analytic approaches, such as mean fields, perturbation theory 
and even renormalization group techniques are also available~\cite{Goldenfeld}. Moreover, the path integral formulation of 
stochastic processes is quite useful to compute correlation and response functions~\cite{WioBook2013}. Even though path integrals for  
multiplicative noise processes have been studied for a long time~\cite{Janssen-RG}, 
important progresses  have been recently reached~\cite{AronLeticia2010, arenas2010, arenas2012, Arenas2012-2, Miguel2015, ArBaCuZoGus2016}. 
Interestingly, this technique provides a useful and beautiful connection between a classical stochastic process and a quantum 
mechanical problem~\cite{Zinn-Justin}.

Preparing the system  at a certain time $t_i$ with a  probability density distribution $P_0(x)$, the solution of the Fokker-Planck 
equation provides a time-dependent probability distribution, $P(x,t)$, at any time $t>t_i$. Equivalently, it can be written in the form
\begin{equation}
P(x,t)=\int  P(x,t|x_i,t_i) P_0(x_i) \; dx_i \; , 
\label{eq:convolution}
\end{equation} 
where $P(x,t|x_i,t_i)$ is the {\em conditional  probability distribution}  of the variable  $x$ at a time $t$, provided 
it was  $x_i$ at a previous time $t_i$.

The conditional probability is a central object in the theory of stochastic processes since  it contains detailed information about  dynamics. 
For instance, equilibrium properties, such as detailed balance,  can be cast in terms of the conditional probability and its time reversal.  
A careful definition of the  time reversal conditional probability, detailed-balance relations, as well as microscopic reversibility  
in multiplicative processes was developed in Ref.~\cite{Arenas2012-2}.
 
 The explicit computation of the  conditional probability  is quite involved. 
One of the methods to compute it  is based in its path integral representation.  For an additive noise Langevin equation, 
the problem is equivalent to the computation of a propagator of a quantum particle. In this context, perturbation theory  
or  semi-classical expansions~\cite{Caroli1979,Caroli1981} can be implemented.  

In this paper, we propose a method to analytically compute a weak noise expansion of the conditional probability in the case of a 
multiplicative process driven by a Langevin equation. We show that, when written in its path integral representation, its computation 
is equivalent to compute a propagator of a quantum particle with variable mass. The state-dependent diffusion is 
translated to a 
quantum equivalent model as a  space-dependent mass.  The quantization of a classical model with variable mass is not uniquely 
defined~\cite{Epele1988}, since the kinetic term is a product of non-commuting (coordinate and momentum) operators. This ambiguity resembles, 
in the stochastic process,   different discretization  prescriptions needed to properly define the stochastic differential equation.
We describe a method to compute a weak noise expansion of the conditional probability in an arbitrary stochastic prescription. 
The key point, in order to compute fluctuations, is the  introduction of a  reparametrized time that transforms the structure of the 
fluctuation operator with multiplicative noise into a simpler form, typically displayed in additive noise problems. In some sense, we are 
transforming multiplicative noise into additive noise by means of a time reparametrization.  This is different from the usual way to 
transform multiplicative into additive noise, by means of a nonlinear variable transformation~\cite{gardiner,vanKampen,Ito1951}.  
Using this technique, fluctuations can be explicitly computed  in the reparametrized time. In the end, it is necessary to invert 
the time transformation in order to have sensible results. 

To illustrate the technique we show a simple nontrivial example. We compute 
the conditional probability of an overdamped harmonic oscillator in the presence of a (nonlinear) multiplicative noise. We obtain
analytic results, allowing to explore the behavior of the conditional probabilities in different time regimes.
Ir order to check the accuracy of the method we compare our results with numerical solution of the Fokker-Planck equation.        

The paper is organized as follow:  in section~\ref{sec:model}, we describe the model and set up the basic notation we use all along the paper. 
In \S~\ref{sec:pathintegral} we show that the conditional probability  of a multiplicative stochastic process is equivalent 
to the propagator on a quantum particle with position-dependent mass. In section~\ref{sec:Weaknoise}, we describe a weak noise expansion 
or semi-classical approximation. Here, we introduce the time reparametrization that allows us to compute fluctuations. 
Finally, in \S~\ref{sec:Oscillator} we show a particular example to illustrate how the computational technique is applied. 
We discuss our results in \S~\ref{sec:discussion}. Additionally, we show details of the calculation in two appendices. 

\section{Langevin description of a multiplicative white noise stochastic evolution}
\label{sec:model}
In order to describe the model and to set up  notations, we briefly review in this section the Langevin and Fokker-Planck description 
of the dynamical evolution of a  multiplicative white noise stochastic process.   

We consider a single random variable $x(t)$ satisfying a first order stochastic differential equation given by
\begin{equation}
dx(t) = f(x(t))dt + g(x(t))dW(t),
\label{eq:Langevin} 
\end{equation}
where $dW=\zeta(t)dt$ is a Wiener process. Thus, $\zeta(t)$ obeys a Gaussian white noise distribution with
\begin{equation}
 \left\langle \zeta(t)\right\rangle   = 0 \;\;\mbox{,}\;\;\;  \left\langle \zeta(t)\zeta(t')\right\rangle = \sigma^2 \delta(t-t')\; .
\label{eq:whitenoise}
\end{equation}
$\sigma$ measures the noise intensity while
$f(x)$ and $g(x)$ are arbitrary smooth functions of $x$, representing the drift force and  the square root of the diffusion function, respectively.  
To completely define equation~(\ref{eq:Langevin}) we fix  the {\em generalized Stratonovich}~\cite{Hanggi1978} prescription 
(also known as $\alpha-$prescription~\cite{Janssen-RG}) to define the stochastic integrals. In this prescription, the integrals are defined 
by the discretization rule
\begin{equation}
\int   g(x(t))\;  dW(t)=\lim _{n\to\infty} \sum_{j=1}^n  g(x(\tau_j))(W(t_{j+1})-W(t_j))
\label{eq:Wiener}
\end{equation}
where $\tau_j$ is chosen from 
\begin{equation}
g(x(\tau_j))=g((1-\alpha)x(t_j)+\alpha x(t_{j+1}))\mbox{~~ with~~} 0\le \alpha \le 1. 
\label{eq:prescription}
\end{equation}
The limit in Eq.~(\ref{eq:Wiener}) is  understood  in the sense of \emph{mean-square limit}~\cite{gardiner}.
The solutions of Eq.~(\ref{eq:Langevin}) depends on the particular value of  $0\le\alpha\le 1$.
Two popular choices are $\alpha=0$, that corresponds with the pre-point It\^o interpretation, and $\alpha=1/2$, with the 
(midpoint) Stratonovich one. Different values of $\alpha$ also imply different calculus rules. The most striking property is the chain rule. 
For an arbitrary function  of the stochastic variable, $Y(x(t))$,  it takes the form~\cite{arenas2012,Arenas2012-2},  
\begin{equation}
\frac{dY(x(t))}{dt}=\frac{\partial Y}{\partial x}\; \frac{dx}{dt} +\frac{(1-2\alpha)}{2}\sigma^2\; \frac{\partial^2 Y}{\partial x^2}\; g^2.
\label{eq.chainrule}
\end{equation} 
As a consequence, integration by parts should be performed with great care. 
Clearly, for $\alpha=1/2$, equation~(\ref{eq.chainrule}) is the usual chain rule. In fact, the Stratonovich prescription is the only one 
in which all the usual calculus rules are preserved.  Sometimes, it is useful to represent the \emph{same stochastic process} in different 
prescriptions. For instance, 
if the process is defined by Eq.~(\ref{eq:Langevin}) in the $\alpha$ prescription, \emph{the same process} can be formulated 
in another $\beta$ prescription by shifting the drift force as 
\begin{equation}
f(x)\to f(x)+(\alpha-\beta) \sigma^2 g(x) g'(x) \; , 
\label{eq:Translator}
\end{equation}
where $g'(x)=dg/dx$, (a rigorous demonstration can be found in the appendix of Ref.~\cite{Miguel2015}),
Using this relation, it  is sometimes useful to consider, instead of  Eq.~(\ref{eq:Langevin}) defined in the $\alpha$ prescription, 
the alternative equation
\begin{equation}
dx(t) = F_S(x(t))dt + g(x(t))dW(t),
\label{eq:LangevinS} 
\end{equation}
with
\begin{equation}
F_S(x)=f(x)+\left(\frac{2\alpha-1}{2}\right)\sigma^2 g(x)g'(x)
\label{eq:FS}
\end{equation}
where, now, the discretization is taken in the Stratonovich prescription.
In this way, solutions  of Eq.~(\ref{eq:LangevinS}), solved in the Stratonovich convention, coincide with those of  Eq.~(\ref{eq:Langevin}), 
solved in the $\alpha$-prescription (with arbitrary $0\le\alpha \le 1$).

Eq.~(\ref{eq:Langevin}) or, equivalently, Eq.~(\ref{eq:LangevinS}), leads~\cite{Lubensky2007} to the
Fokker-Planck equation
\begin{equation}
\frac{\partial P(x,t)}{\partial t}+\frac{\partial J(x,t)}{\partial x}=0,
\label{eq:Continuity}
\end{equation} 
where  $P(x,t)$ is the time-dependent probability distribution and  the probability current is  given by
\begin{align}
J(x,t)&= \left[f(x)-(1-\alpha) g(x)g'(x)\right] P(x,t)\nonumber \\
&-\frac{1}{2}g^2(x) \frac{\partial P(x,t)}{\partial x}.
\label{eq:J(x,t)}
\end{align} 
This equation has a time independent equilibrium solution $P_{\rm eq}(x)$ which satisfies $J(x)=0$.  It has the form 
\begin{equation}
P_{\rm eq}(x)={\cal N}\; e^{-\frac{1}{\sigma^2}U_{\rm eq}(x)},
\label{eq:Peq}
\end{equation} 
where ${\cal N}$ is a normalization constant and 
\begin{equation}
U_{\rm eq}(x)= -2\int^x \frac{f(x')}{g^2(x')}dx'+(1-\alpha)\sigma^2\ln g^2(x).
\label{eq:U}
\end{equation} 

In most practical cases, in the absence of noise, the system is conservative, characterized by a potential  $U(x)$. In these cases, the relation between the drift force and the potential is
\begin{equation}
f(x)=-\frac{1}{2}g^2(x) \frac{dU(x)}{dx}\; , 
\label{eq:Einstein}
\end{equation}
that can be considered as a generalization of the Einstein relation for Brownian motion to the case of multiplicative noise~\cite{Arenas2012-2}. 
Replacing Eq.~(\ref{eq:Einstein}) into Eq.~(\ref{eq:U}) we find for the equilibrium potential
\begin{equation}
U_{\rm eq}(x)= U(x)+2(1-\alpha)\sigma^2\ln g(x)\; .
\label{eq:Ueq}
\end{equation} 
The equilibrium distribution depends not only on the given functions  $(U(x),g(x))$,  but also on the value of the $\alpha$-prescription 
which defines the Wiener integral. The only prescription that leads to the Boltzmann distribution $U_{\rm eq}(x)=U(x)$ is $\alpha=1$; for this reason, 
this convention is sometimes called ``thermal
prescription'' or even H\"anggi-Klimontovich interpretation~\cite{Hanggi1982,Klimontovich}. Furthermore, this prescription is also known as 
anti-It\^o  interpretation. Interestingly, it can be considered as the time reversal conjugated to 
the It\^o prescription~\cite{Arenas2012-2,Miguel2015}.

Choosing convenient values for $\alpha$ and $U(x)$, it is possible to study,
in a unified formalism, model systems with general equilibrium distributions that
go from Boltzmann thermal equilibrium 
to power-law distributions. A simple example of the latter case is to consider
a pure noisy system with $U(x)=0$. In that case, the equilibrium distribution
is $P_{\rm eq}\sim g(x)^{-2(1-\alpha)}$.

\section{Classical stochastic evolution and quantum mechanics of a particle with variable mass }
\label{sec:pathintegral}
The transition probability $P(x_f, t_f|x_i, t_i)$ is a central ingredient in the study of any dynamical property of a stochastic process.
It  represents the conditional probability of finding the system in the state $x_f$ at time $t_f$, provided the
system was in the state $x_i$ at time $t_i$. In the path integral formalism, it
can be written as~\cite{arenas2010}
\begin{equation}
P(x_f, t_f|x_i, t_i)=
\int {\cal D}x\; {\det}^{-1}(g)\;  e^{-\frac{1}{\sigma^2}S[x]} \; ,
\label{eq:Pfw}
\end{equation}
where the ``action'' $S[x]$ is given by
\begin{equation}
 S[x] = \int_{t_i}^{t_f} dt \left\lbrace\frac{1}{2g^2} \left[ \frac{dx}{dt} - F_S +
\frac{1}{2} \sigma^2 g g'\right]^2+\frac{\sigma^2}{2}   F'_S \right\rbrace ,
\label{eq:S0}
\end{equation}
with boundary conditions $x(t_i)=x_i$ and $x(t_f)=x_f$. 
Eqs.~(\ref{eq:Pfw}) and~(\ref{eq:S0}) are the Onsager-Mashlup representation~\cite{Onsager1953} of the transition probability of the 
stochastic process driven by Eq.~(\ref{eq:LangevinS}).  Note that, for simplicity,  we have built  a path integral representation of 
Eq.~(\ref{eq:LangevinS}). In this way, we can employ usual calculus rules to deal with the path integral. The information about the different 
stochastic prescriptions is codified in $F_S$, given by Eq.~(\ref{eq:FS}).  Alternatively, we could choose to represent Eq.~(\ref{eq:Langevin}) 
in the past integral formalism; 
in this case, it would be necessary to use the $\alpha$-generalization of stochastic calculus~\cite{ Arenas2012-2,Miguel2015}. Of course, 
the results are exactly the same, the chosen representation is just a matter of convenience.    
 
 It is instructive to rewrite the action in an alternative way.
Expanding the squared bracket in Eq.~(\ref{eq:S0}),  using Eqs.~(\ref{eq:FS}) and~(\ref{eq:Einstein}) and integrating by parts,  we find 
\begin{equation}
 S[x] = \frac{\Delta U_{\rm eq}}{2}+\int_{t_i}^{t_f} dt \;  L(x,\dot x) \; .
 \label{eq:S}
\end{equation}
Here,  the first term is a state function governed by the equilibrium potential $U_{\rm eq}$ evaluated at  the initial and the final state of the system,  
$\Delta U_{\rm eq}=U_{\rm eq}(x_f)-U_{\rm eq}(x_i)$, with  $U_{\rm eq}$  given by Eq.~(\ref{eq:Ueq}). 
The Lagrangian can be written in the suggestive form, 
\begin{equation}
 L= \frac{1}{2}\left(\frac{1}{g^2(x)}\right) \dot x^2+V(x) \; ,
\label{eq:L}
\end{equation}
where
\begin{equation}
  V(x) = \frac{g^2}{2}\left[\left(\frac{U'_{\rm eq}}{2}\right)^2 - \sigma^2\left(\frac{U''_{\rm eq}}{2}+\frac{g'}{g} U'_{\rm eq} \right) \right]
  + \frac{\sigma^4}{4}\left(g g'\right)' .
   \label{eq:V}
\end{equation}
The primes, $(~)'$, means derivative with respect to $x$.
 Replacing Eq.~(\ref{eq:S}) into Eq.~(\ref{eq:Pfw}),  the conditional probability  takes the form
\begin{equation}
P(x_f, t_f|x_i, t_i)= e^{-\frac{\Delta U_{\rm eq}}{2\sigma^2}} K(x_f, t_f|x_i, t_i)
\label{eq:PK}
\end{equation}
where the \emph{propagator} $K(x_f, t_f|x_i, t_i)$ is given by 
\begin{equation}
K(x_f, t_f|x_i, t_i)=\int [{\cal D}x]\;  e^{-\frac{1}{\sigma^2}\int_{t_i}^{t_f} dt  \; L(x,\dot x)} \; .
\label{eq:Propagator}
\end{equation}
Here, the functional integration measure is
\begin{equation}
[{\cal D}x]={\cal D}x\;{\det}^{-1} g=\lim\limits_{\substack{{N\to\infty} \\ {\Delta t\to 0}}} \prod_{n=0}^N \frac{dx_n}{\sqrt{\Delta t \;g^2(\frac{x_n+x_{n+1}}{2})}}
\label{eq:Dx}
\end{equation}
where $x_n=x(t_n)$, with $x(t_0)=x_i$ and $x(t_N)=x_f$. 

Interestingly, Eq.~(\ref{eq:Propagator}) is the exact propagator of a quantum particle with position-dependent mass $m(x)=1/g^2(x)$ 
moving in a potential $V(x)$, written in the imaginary time path integral formalism $t\to -it$.  The noise $\sigma^2$ places the role of 
$\hbar$ in the quantum theory.  There is subtlety in the identification of the stochastic and the quantum problem. Inside $V(x)$, 
Eq.~(\ref{eq:V}), there are terms of order $\sigma^2$ and $\sigma^4$. These terms are absent in a pure quantum problem, where the potential 
is independent of $\hbar$. However, as we will show in next section, this fact is important in  developing  a weak noise (or semi-classical) 
approximation since the saddle-point approximation has contributions of the same order than fluctuations.   

The quantization of a classical system with variable mass has ambiguities due to operator ordering. In fact, 
the quantum kinetic term mixes the non-commuting operators, position $\hat x$ and momentum $\hat p$. As a consequence, 
different orderings in the Hamiltonian are not equivalent. In the configuration space path integral approach, this fact is reflected 
in different discretization schemes of the time integral. Then, the quantum ordering problem is associated in the classical Langevin 
description with the different stochastic prescriptions available to compute the Wiener integrals. Specifically, the propagator of 
Eq.~(\ref{eq:Propagator}) with time integrals symmetrically discretized (Stratonovich  prescription $\alpha=1/2$),  
is equivalent to a quantum problem described by  a Hamiltonian operator with the Weyl order~\cite{Epele1988}, 
\begin{equation}
\hat H= \frac{1}{4}\left( g^2(\hat x)\hat p^2 + 2 \hat p \;  g^2(\hat x)\;  \hat p+\hat p^2 g^2(\hat x)\right)+V(\hat x) .
\end{equation}
It is worth mentioning that, in our formalism, the dependence of the conditional probability on the stochastic prescription 
$\alpha$ is completely taken into account through the expression of the equilibrium potential $U_{\rm eq}(x)$, given by Eq.~(\ref{eq:Ueq}).   
 
In the next section, we present a detailed procedure for analytically computing the propagator of Eq.~(\ref{eq:Propagator}) in a weak noise expansion. 

\section{Weak noise expansion}
\label{sec:Weaknoise}
In order to explicitly compute the conditional probability, we  make a weak noise expansion, in some sense,  equivalent to a semi-classical 
(or WKB) approximation  in quantum mechanics. The general method is very well known in additive noise processes and in quantum systems.
The main idea is that the functional integral of Eq.~(\ref{eq:Pfw}) is dominated, at very low noise,  by  extrema of the action, 
in such a way that at first approximation 
\begin{equation}
P(x_f, t_f|x_i, t_i)\sim \sum_j e^{-\frac{1}{\sigma^2}S[x^{(j)}_{cl}]} \;\; ,
\label{eq:Psp}
\end{equation}
where $x^{(j)}_{cl}$ (with $j=1,2,\ldots$) are different solutions of the classical equation of motion with the same boundary conditions.
To improve the approximation, the functional integral of Eq.~(\ref{eq:Pfw}) could be computed considering small fluctuations around each solution
$x^{(j)}_{cl}$. This integration produces a prefactor in Eq.~(\ref{eq:Psp}) given in terms of a functional determinant (see below)  
that, upon exponentiation, will result in an effective action containing terms proportional to $\sigma^2$ (or $\hbar$ in a quantum system).
 The application of this procedure to multiplicative stochastic processes or, equivalently,  to variable mass quantum problems, 
is cumbersome~\cite{Langouche1979}. In particular, we will see that the classical solution as well as fluctuation contributions 
to the effective action, both have $\sigma$-dependent terms. Thus, it will be necessary to carefully compute and consider both contributions 
in order to get a consistent approximation.

In this section, we present a general procedure  for the computation of the conditional probability 
in a stochastic process driven by a  drift force $f(x)$ and a diffusion function $g(x)$, for a general stochastic prescription $\alpha$. 

We begin by assuming that, at very weak noise, the functional integral, Eq.~(\ref{eq:Pfw}) is dominated by the solutions of  the  
classical equation  $\delta S[x]/\delta x(t)=0$,
where $S[x]$ is given by Eq.~(\ref{eq:S}). Explicitly computing the functional derivative, we find 
 \begin{equation}
 \frac{d^2 x}{dt^2}=g^2 V' +\frac{g'}{g}  \dot x^2  \;. 
 \label{eq:SadlePoint}
 \end{equation}
This equation provides an alternative interpretation of the classical problem. It corresponds with the classical dynamics of a unit mass particle,
 moving under the influence of a non-conservative velocity dependent force, proportional to $\dot x^2$~\cite{Epele1988}. This term is proper 
 of multiplicative systems since it is proportional to $g'(x)$.
 The solution of Eq.~(\ref{eq:SadlePoint}) is quite involved due, on one hand, to the nonlinearity introduced by the diffusion function  $g(x)$ and, 
  on the other, to the nonlinearity in the velocity. However,  using the fact that the Lagrangian is invariant under time translations, 
  it is  not difficult  to built up a first integral of this equation. In fact, the canonical momentum 
  $p=\partial {L}/\partial \dot x= \dot x/g^2(x)$ and   the classical Hamiltonian  $H(x,p)=\dot x p- L$ can be simply computed, obtaining
\begin{equation}
H(x,p)=\frac{1}{2} g^2(x) p^2 - V(x) \;.
\label{eq:H}
\end{equation}
This is a classical Hamiltonian, thus, there is no ordering problem of the kinetic term. 
The Hamiltonian, evaluated on a solution of the equation of motion $x_{cl}(t)$
\begin{equation}
H(x_{cl}(t))=\frac{1}{2} \frac{1}{g^2(x_{cl})}\dot x_{cl}^2  - V(x_{cl})  
\label{eq:Hcl}
\end{equation}
should be  a conserved quantity,  $dH/dt=0$. Then, 
\begin{equation}
\dot x_{cl}^2=2 g^2_{cl}\left(V_{cl}+H\right) \, , 
\label{eq:vsquare}
\end{equation}
where $g_{cl}=g(x_{cl}(t))$ and $V_{cl}=V(x_{cl}(t))$.
Note that the Hamiltonian, even though it is a conserved quantity, is  not the energy, since the Lagrangian is not a 
homogeneous quadratic function of the velocity.  

From Eq.~(\ref{eq:vsquare}), we can write  the solution of Eq.~(\ref{eq:SadlePoint}) by a quadrature
\begin{equation}
t-t_0= \int_0^{x_{cl}}  \frac{ds}{\sqrt{2  V_{\rm eff}(s)}}\; , 
\label{eq:firstIntegral}
\end{equation} 
where we have defined an effective potential 
\begin{equation}
V_{\rm eff}(x)=  g^2(x)\left[V(x)+H\right].
\label{eq:Veff}
\end{equation}
These expressions have two arbitrary constants, $t_0$ and $H$, that should be determined by means of  the boundary 
conditions $x_{cl}(t_i)=x_i$ and $x_{cl}(t_f)=x_f$. Thus, Eqs.~(\ref{eq:firstIntegral}) and~(\ref{eq:Veff})  implicitly define $x_{cl}(t)$, 
used as a starting point of the weak noise approximation. As anticipated,  due to the fact that $V(x)$ contains terms proportional to $\sigma^2$ (see Eq. (\ref{eq:V})), $S[x_{cl}]$ will contain such terms. Therefore,  for consistency,  we need to go beyond the saddle-point approximation. The following subsection is devoted to this matter. 

\subsection{Reparametrization of time and Gaussian fluctuations}
Let us assume, for the moment, that, given initial and final conditions, the classical solution $x_{cl}$ is unique. 
The genera\-li\-za\-tion to multiple solutions is straightforward. Then, we consider fluctuations around it   
\begin{equation}
x(t)=x_{cl}(t)+\delta x(t) \; , 
\end{equation}
with boundary conditions $\delta x(t_i)=\delta x(t_f)=0$\; .
Replacing the expansion into Eq.~(\ref{eq:Propagator}) and keeping up to second-order terms in the fluctuations we find for the propagator
\begin{align}
K(x_f, t_f|x_i, t_i)&=e^{-\frac{1}{\sigma^2}S_{cl}} 
\label{eq:PropagatorF} \\
&\times \int [{\cal D}\delta x]\; 
e^{-\frac{1}{2}\int dtdt'\; \delta x(t) O(t,t') \delta x(t') }    \; ,
\nonumber
\end{align}
where the classical action $S_{cl}$ is
\begin{equation}
S_{cl}=\int_{t_i}^{t_f} dt  \; L(x_{cl}(t),\dot x_{cl}(t))
\label{eq:Scl}
\end{equation}
and the fluctuation kernel
\begin{equation}
O(t,t')=-\frac{d~}{dt}\left( \frac{1}{g^2_{cl}}\frac{d\delta(t-t')}{dt}\right)+\left(\frac{1}{g^2_{cl}}V'_{\rm eff}(x_{cl})\right)'\delta(t-t)\; .
\label{eq:kernel}
\end{equation}
In Eq.~(\ref{eq:PropagatorF}),  the functional integration measure is 
\begin{equation}
[{\cal D}\delta x]=\lim\limits_{\substack{{N\to\infty} \\ {\Delta t\to 0}}} \prod_{n=0}^N \frac{d\delta x_n}{\sqrt{\Delta t \;g^2(\frac{x_{cl}(t_n)+x_{cl}(t_{n+1})}{2})}} \; .
\label{eq:Ddeltax}
\end{equation}
Although Eq.~(\ref{eq:PropagatorF}) formally looks like a Gaussian integral, its evaluation is not simple. 
The reason is twofold. On one hand, the fluctuation kernel $O(t,t')$, Eq.~(\ref{eq:kernel}), is not trivial due to the time dependence of 
$g_{cl}=g(x_{cl}(t))$. On the other, the integration measure, Eq.~(\ref{eq:Ddeltax}),  has the diffusion function $g(x(t))$ in the denominator.  
This factor comes from  $\det^{-1}g$ in Eq.~(\ref{eq:Pfw}) and resembles  curvature effects in the time axes. 

 In order to compute the fluctuation integral, we make a time  reparametrization.
 For concreteness, we introduce a new time variable $\tau$, by means of 
\begin{equation}
\tau=\int_0^t g^2(x_{cl}(t')) dt'\; .
\label{eq:reparametrization}
\end{equation}
This is a nontrivial \emph{local} scale transformation, weighted by the diffusion function, evaluated at the classical solution $x_{cl}(t)$. 
Eq.~(\ref{eq:reparametrization}) defines a function $\tau(t)$, in such a way that
\begin{equation}
\frac{d\tau}{dt}=g^2(x_{cl}(t))\; . 
\end{equation}
We have chosen the integration constant in order to keep the time origin unchanged, $\tau(0)=0$.

Performing this time reparametrization,  the fluctuation kernel transforms as $O(t,t')\to \Sigma(\tau,\tau')$, and takes the simpler form 
\begin{equation}
\Sigma(\tau,\tau')=\left[-\frac{d^2~~}{d\tau^2}+W[x_{cl}]\right]\delta(\tau-\tau')
\label{eq:Sigma}
\end{equation}
where 
\begin{equation}
W(x_{cl})=
\frac{1}{g^2_{cl}}\left(\frac{1}{g^2_{cl}}V'_{\rm eff}(x_{cl})\right)'\; .
\label{eq:W}
\end{equation}
On the other hand, after discretizing the reparametrized time axes $\tau$, the functional integration  measure, Eq.~(\ref{eq:Ddeltax}) becomes  
\begin{equation}
[{\cal D}\delta x]=\lim\limits_{\substack{{N\to\infty} \\ {\Delta \tau\to 0}}} \prod_{n=0}^N \frac{d\delta x_n}{\sqrt{\Delta \tau}} \; ,
\label{eq:Ddeltaxtau}
\end{equation}
in which the function $g(x_{cl})$ has been absorbed in the reparametrization. 

Thus, in the new  time variable $\tau$, the functional integral of fluctuations  is Gaussian and can be formally evaluated, 
obtaining for the propagator
\begin{equation}
K(x_f, t_f|x_i, t_i)=\left( \det\Sigma(\tau_i,\tau_f)\right)^{-1/2} e^{-\frac{1}{\sigma^2}S_{cl}(t_i,t_f)}
\label{eq:PropagatorWeaknoise}
\end{equation}
where the relation between $(\tau_i,\tau_f)$ and $(t_i,t_f)$ is given through Eq.~(\ref{eq:reparametrization}). 

 Considering that Eq.~(\ref{eq:PropagatorWeaknoise}) is the main result of this section, let us comment on its meaning.  
The general formal structure of the propagator is the usual one in a semiclassical approximation. It consists on an exponential 
of a classical action and a prefactor which codifies fluctuations. However, in the present case of multiplicative noise processes, 
there are important differences that we should stress.  One of the effects of the multiplicative noise in the classical action 
is to correct terms proportional to  $\sigma^2$ and  to produce new terms of order $\sigma^4$, as can be seen from the definition of 
the potential $V(x)$,  Eq.~(\ref{eq:V}). Thus, in order to correctly implement the appro\-xi\-ma\-tion scheme, it is essential to compute fluctuations, 
since they will contribute with terms of the same order.   
The explicit computation of the prefactor is very interesting. The transformed fluctuation operator $\Sigma(\tau,\tau')$, Eq.~(\ref{eq:Sigma}), 
has exactly the same structure as the fluctuation operator of an additive process, albeit with a modified potential $W(x)$, Eq.~(\ref{eq:W}). 
This fact allows us to compute the functional determinant using the usual  techniques developed to treat additive noise. However, 
the result of this computation is expressed in a reparametrized time variable $\tau$. In order to produce sensible results, at the end of 
the calculations,  we need to go back to real time, $t$, by inverting the reparametrization transformation.
Summarizing, we have transformed the computation of  fluctuations in a multiplicative stochastic process into a simpler problem of an additive 
process by means of a  time reparametrization.

\section{Transition probability for an overdamped harmonic oscillator with multiplicative noise}
\label{sec:Oscillator}
In this section, we illustrate the method proposed in this paper by analyzing the simplest nontrivial example of a  multiplicative stochastic process. 
Let us consider an overdamped harmonic oscillator in the presence of multiplicative noise. A lot of particular examples of this system have been
ana\-ly\-zed for different physical applications. A good peda\-go\-gi\-cal presentation with lots of references  can be found in 
Ref.~\onlinecite{Gitterman}.  
Much of these references try to compute the Fokker-Planck equation or some moments of the probability distribution. 
Differently, we are interested here in the transition probability with the main purpose of explicitly showing how the technique 
described in this paper (specially the time reparametrization) works.     

For concreteness, let us choose a harmonic oscillator potential    
\begin{equation}
U(x)=\frac{\omega}{2} x^2 \, ,
\label{eq:UOs}
\end{equation}
where $\omega$ is the natural frequency of the oscillator and
a diffusion function
\begin{equation}
g(x)=1+\lambda^2 x^2 ,
\label{eq:gOs}
\end{equation}
where $\lambda$ controls the intensity of the multiplicative component of the  noise. In particular, for $\lambda=0$, we recover 
the usual additive white noise process.

Using the  generalized Einstein relation, Eq.~(\ref{eq:Einstein}), the drift force is
\begin{equation}
f(x)=-\frac{\omega}{2} x \left(1+\lambda^2 x^2\right)^2 .
\end{equation}
This relation guarantees that the equilibrium potential, Eq.~(\ref{eq:Ueq}), is given by 
\begin{equation}
U_{\rm eq}(x)=\frac{\omega}{2} x^2+2(1-\alpha)\sigma^2\ln\left(1+\lambda^2 x^2\right) \;, 
\label{eq:UeqOs}
\end{equation} 
implying, for the asymptotic equilibrium probability distribution, 
\begin{equation}
P_{eq}(x)={\cal N }\frac{e^{-\frac{\omega}{2\sigma^2} x^2}}{\left(1+\lambda x^2\right)^{2(1-\alpha)}} 
\label{eq:PeqOs}
\end{equation}
where ${\cal N}$ is a normalization constant.
One of the effects of multiplicative noise is to correct the Gaussian equilibrium distribution with  a power law factor which depends on the 
stochastic discretization prescription $\alpha$. 
 We also observe that, for the H\"anggi-Klimontovich prescription ($\alpha=1$), the equilibrium probability distribution is of Boltzmann type, 
 completely equivalent to a harmonic oscillator with additive noise. 

We want to compute the propagator $K(x,T|0,0)$ for this model in the weak noise approximation by using Eq.~(\ref{eq:PropagatorWeaknoise}),  
where we have chosen initial conditions
$t_i=0$, $x_i=0$, and final conditions $t_f=T$, $x_f=x$.

The first step in the weak noise approximation is to compute the classical action $S_{cl}$. 
To do this, we need to compute the solution of the saddle-point equation, Eq.~(\ref{eq:SadlePoint}) or, equivalently, Eq.~(\ref{eq:firstIntegral}), 
and to replace it into Eq.~(\ref{eq:Scl}).
We have computed $S_{cl}$ considering a weak multiplicative effect, $|\lambda x|<<1$. That is, the results are accurate in a distance 
range, $x^2<<1/\lambda^2$, where the diffusion function $g(x)$ is not very far from its additive value $g=1$. 
For details of the calculations, Appendix~\ref{Ap:SP} can be consulted. We have obtained 
\begin{align}
 S_{cl}=&\frac{\omega x^2}{4} \coth(\omega T/2)-\sigma^2\frac{\omega T}{4}
 \nonumber \\
 &-\frac{\sigma^2 \lambda^2 x^2(1+2\alpha)}{2}\left(1-\frac{\omega T}{\sinh(\omega T)}\right) \coth(\omega T/2)
 \nonumber \\
 &+O(\lambda^4 x^4)\; .
  \label{eq:SclOs}
 \end{align} 
The first line of this expression  is the usual result for the harmonic oscillator with additive noise. The second line is the 
correction due to the multiplicative noise to order $\lambda^2 x^2$. For very weak noise, only the first term of Eq.~(\ref{eq:SclOs}) is relevant.  
An important observation is that we are getting terms of order $\sigma^2$  at  saddle-point level. On the other hand, 
fluctuations will contribute also with $\sigma^2$ terms. Consequently, this approximation is only consistent provided 
fluctuations are taken into account. 

As described  in the preceding section, to  properly compute fluctuations
we need to reparametrize  the time variable according to Eq.~(\ref{eq:reparametrization}). In the present example, this reparametrization reads 
\begin{equation}
\tau=t\left\{1+\frac{\lambda^2 x^2}{\sinh^2(\omega T/2)}\left(\frac{\sinh\omega t}{\omega t}-1\right) \right\}
\label{eq:tau-t}
\end{equation}
where $0<t< T$.
With this reparametrization the computation of fluctuations reduces to evaluating the determinant of the operator  of Eq.~(\ref{eq:Sigma}), 
where we have, now,
 \begin{equation}
 W(x_{cl})=\frac{\omega^2}{4}+\frac{\lambda^2 x^2\omega^2}{2\sinh^2(\omega T/2)} \left(1+8 \sinh^2(\omega \tau/2)\right) \;.
 \label{eq:Wosc1}
 \end{equation}
 In Appendix~\ref{Ap:fluctuations}, we explicitly compute this determinant in the reparametrized time axes and, after that, 
 we turn back to ori\-gi\-nal  time variable $T$.
The result at order $\lambda^2 x^2$ is 
 \begin{widetext}
\begin{align}
 &\frac{1}{\sqrt{\det\Sigma(0,\tau)}}=\sqrt{\frac{\omega T}{2\sinh(\omega T/2)}}\left\{ 1- \frac{\lambda^2x^2}{2\sinh^2(\omega T/2)}  
 \left[\frac{3}{2}+\frac{1}{2}\frac{\sinh(3\omega T/2)}{\sinh(\omega T/2)}+ \frac{\omega T}{2}  \mbox{coth}(\omega T/2)\left(\frac{\sinh(\omega T)}{\omega T}-4\right)
  \right]\right\} \; .
  \label{eq:detsResult}
 \end{align}
\end{widetext}

 Since Eq~(\ref{eq:detsResult}) is an expansion in  $(\lambda x)^2$, we can exponentiate the second term and absorb it in the  
 definition of an effective action, finally obtaining for the propagator
 \begin{eqnarray}
K(x, T|0,0)= \sqrt{\frac{\omega T e^{\omega T/2 }} {2\sinh(\omega T/2)}} e^{-\frac{1}{\sigma^2}S_{\rm eff}}
\label{eq:PropagatorOsc}
\end{eqnarray}
where
\begin{widetext}
\begin{equation}
S_{\rm eff}=\frac{\omega x^2}{4}\mbox{coth}(\omega T/2)\left\{1+ \frac{\lambda^2\sigma^2  }{\omega}
 \left[ 4(1-\alpha)+  \frac{2}{\sinh(\omega T)}\left( 5+3 e^{-\omega T}+\omega T\left[ 1+2\alpha-4{\rm coth}(\omega T/2) \right]  \right)  \right]
 \right\} .
\label{eq:Seff}
\end{equation}
\end{widetext}

This is the central result of this section. We have  computed, in  closed analytic form, the propagator for a harmonic oscillator under 
the influence of a nonlinear multiplicative noise in the weak noise approximation.  
The first term of Eq.~(\ref{eq:Seff}) is the usual result for a harmonic oscillator under additive noise, while the second term 
represents the corrections due to the multiplicative character of the noise.

It is instructive to analyze particular limits in the time scale.  For instance, consider $\omega T<<1$. In this case, Eq.~(\ref{eq:Seff}) 
takes the simpler form
\begin{equation}
S_{\rm eff}=\frac{1}{2}\frac{x^2}{T}\left\{1+\frac{5}{3}\left(\frac{\lambda^2\sigma^2}{\omega}\right)\omega T+\mbox{O}[(\omega T)^2]\right\} \; .
 \label{eq:Tsmall}
 \end{equation}
The first term is the result for a diffusive free particle.  This is, indeed, the correct result since, for very short times, 
the particle did not have enough time to explore the harmonic potential. The first correction in $\omega T$ is due to multiplicative noise 
(order $\lambda^2$) and is independent of the stochastic prescription; $\alpha$-dependent terms only appear at order $(\omega T)^2$. 
 
 On the other hand, in the asymptotic limit  $\omega T>>1$, the effective action is 
\begin{equation}
S_{\rm eff}=\frac{1}{4} \omega x^2+(1-\alpha)\lambda^2x^2\sigma^2   
+ \mbox{O}(1/\omega T) \; .
 \label{eq:Tbig}
 \end{equation}
Replacing this result into Eq.~(\ref{eq:PropagatorOsc}) we get, for the conditional  probability Eq.~(\ref{eq:PK}),
\begin{equation}
\lim_{\omega T \rightarrow \infty} P(x,T|0,0)=P_{eq}(x)={\cal N}\;  e^{-U_{eq}(x)/\sigma^2} \, ,
\end{equation}
which is the correct result obtained, independently, by solving the stationary Fokker-Planck equation.

\subsection{Comparison with the exact Fokker-Planck solution}
\begin{figure}
\centering
\includegraphics[scale=0.8]{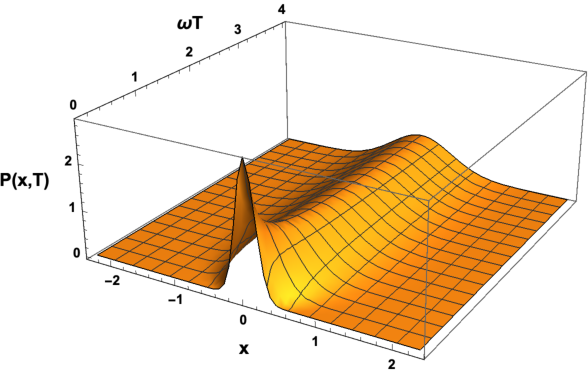}
\caption{Solution of the Fokker-Planck equation, Eqs.~( (\ref{eq:Continuity}) and (\ref{eq:J(x,t)})), for the model given by Eqs.~(\ref{eq:UOs}) 
and~(\ref{eq:gOs}) with initial condition $P(x,0)=\delta(x)$ and the following values of the 
parameters: $\omega=1$, $\lambda=0.1$, $\sigma=0.5$ and $\alpha=0$.}
\label{fig:FP}
\end{figure}

In order to check the method and the  accuracy of our approximation, we compare the analytic result with the numerical solution of the 
Fokker-Planck equation. 
In fact, from Eq.~(\ref{eq:convolution}), it is simple to check that the conditional proba\-bi\-li\-ty $P(x,t|0,0)$ is  the solution of 
the Fokker-Planck equation, $P(x,t)$, with the initial condition $P(x,0)=\delta(x)$. Therefore, we  numerically solved 
equations~(\ref{eq:Continuity}) and~(\ref{eq:J(x,t)}),  for our model Eqs.~(\ref{eq:UOs}) and~(\ref{eq:gOs}). In Figure~\ref{fig:FP}, 
we depict a solution for typical values of the parameters, $\omega=1$, $\lambda=0.1$, $\sigma=0.5$ and $\alpha=0$.  
The initial probability density is  strongly peaked at $x=0$. It   diffuses  in time and,  at long times, saturates  to the exact analytic expression
Eq.~(\ref{eq:PeqOs}).  We observe  that, as already commented, the asymptotic  equilibrium probability depends on the stochastic 
prescription $\alpha$. For anti--It\^o prescription, $\alpha=1$, the probability density is Gaussian. However, for $\alpha\neq 1$, the Gaussian 
behavior is slightly corrected by a power law. In the region $\lambda^2 x^2<<1$, the distribution can be approximated by a Gaussian with variance 
$\Gamma^2=1-4(1-\alpha)\lambda^2\sigma^2/\omega$.  This behavior, deduced from Eq.~(\ref{eq:PeqOs}), is verified by the numerical solution 
of the Fokker-Planck equation with great precision. On one hand, we see that the effect of the multiplicative character of the noise 
is controlled by the parameter $\beta\equiv \lambda^2\sigma^2/\omega$. We also note from Figure~\ref{fig:FP} that, for $\omega T>3$, 
the asymptotic limit is already reached.   
On the other hand, our approximate analytic result, valid for  $\omega \sigma^2<<1$ and $\lambda x<<1$,  can be written 
in the form of a normal distribution, 
\begin{equation}
P(x,T|0,0)= \sqrt{\frac{\omega}{2\pi\sigma^2 \Gamma^2(T)}} \exp\left(-\frac{\omega x^2}{2\sigma^2 \Gamma^2(T)}\right)
\end{equation}
where the dimensionless time-dependent variance is given by
\begin{widetext}
\begin{equation}
\Gamma^2(T)=\left(1-e^{-\omega T}\right)
\left\{1-\frac{\lambda^2\sigma^2}{\omega}\left[4(1-\alpha)+
\frac{e^{-\omega T/2}}{8\sinh(\omega T/2)}\left( 5+3 e^{-\omega T}+\omega T\left[ 1+2\alpha-4{\rm coth}(\omega T/2) \right]   \right)   \right] 
\right\} .
\label{eq:Gamma}
\end{equation}
\end{widetext}
The first term in the last expression corresponds to the usual additive noise result, while the second one, proportional 
to $\beta=\lambda^2\sigma^2/\omega$, is the
multiplicative noise correction. From this equation, it is immediate to verify that in the asymptotic limit $\omega T>>1$, we get 
the correct result  $\Gamma^2=1-4(1-\alpha)\lambda^2\sigma^2/\omega$, while at very short time approximation, $\omega T<<1$, 
we reach the initial distribution $\lim_{\omega T\to 0}P(x,T|0,0)=\delta(x)$. 

\begin{figure}[hbt]
\centering
\includegraphics[scale=0.8]{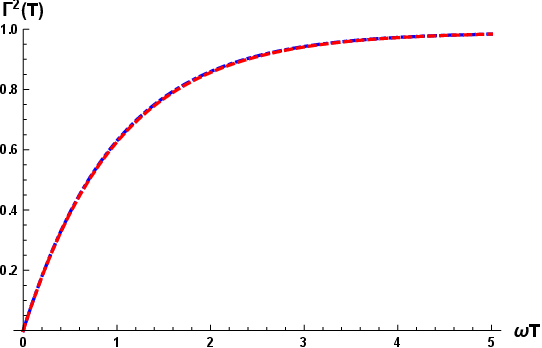}
\caption{$\Gamma^2(T)$ computed from Eq.~(\ref{eq:Gamma}) and from Eq.~(\ref{eq:Gammaexact}) for  $\alpha=0$, $\sigma=0.5$, $\lambda=0.1$. 
Both curves coincide within the graphic precision. }
\label{fig:Gamma2}
\end{figure}

Now, we  compare the approximate analytic solution 
with the exact numerical distribution in all the time range for different values of the parameters. 
For this, we observe that the variance can be obtained from the maximum of the distribution probability  at $x=0$, as 
$\Gamma^2(T)=\omega^2/\sqrt{2\pi \sigma^2 P(0,T|0,0)^2}$. In the same way, we define the quantity 
\begin{equation} 
 \Gamma_{\rm ex}^2(T)=\sqrt{\frac{\omega}{2\pi \sigma^2 P(0,T)^2}} \;,
\label{eq:Gammaexact}
 \end{equation}
  computed from the numerical solution, $P(0,T)$, of the Fokker-Planck equation. 
In Figure~\ref{fig:Gamma2}, we depict both curves computed in the It\^o prescription, $\alpha=0$, with a moderate value of 
the noise $\sigma=0.5$ and $\lambda=0.1$. We see that both curves coincide in the range within the graphic precision. 
Although we are working in the  weak noise approximation, Eq.~(\ref{eq:Gamma}) is quite accurate even for noise intensities as big as $\sigma=1.5$.

\begin{figure}[hbt]
\begin{center}
\subfigure[]
{\label{fig:Gamma4}
\includegraphics[width=0.4\textwidth]{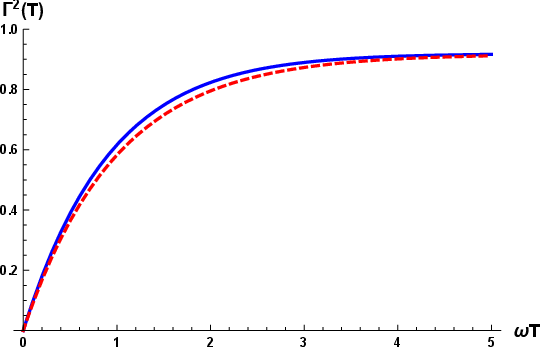}}
\subfigure[]
{\label{fig:Gamma12}
\includegraphics[width=0.4\textwidth]{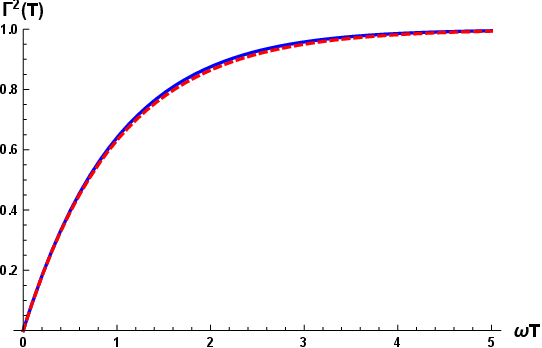}}
\end{center}
\caption{Comparison of the variance $\Gamma^2(T)$, computed with Eq.~(\ref{eq:Gamma}), with the exact numerical evaluation $\Gamma^2_{\rm ex}(T)$. 
The solid line is  $\Gamma^2(T)$, while $\Gamma^2_{\rm ex}(T)$ is depicted by the dotted line. We used the following parameters for both subfigures: $\omega=1$, $\sigma=1.5$, $\lambda=0.1$. Subfigure (a) was computed in the 
It\^o prescription, $\alpha=0$,  while  Subfigure (b) was calculated in the H\"anggi-Klimontovich prescription, $\alpha=1$.}
\label{fig:Gamma4-12}
\end{figure}

In Figure~\ref{fig:Gamma4-12}, we show two curves computed with the same parameters values $\sigma=1.5$, $\lambda=0.1$, but 
for different stochastic prescriptions. 

~\ref{fig:Gamma4} is computed in the It\^o prescription while Subfigure~\ref{fig:Gamma12} is computed in the anti-It\^o interpretation.   
We see that the curves fit quite well for initial values of the time evolution as well as in the asymptotic limit. 
In the intermediate range, we begin to observe a small deviation due to the big value of the noise. It can also be noticed from this figure that a 
better approximation is obtained for the H\"anggi-Klimontovich interpretation ($\alpha=1$). 

\begin{figure}
\centering
\includegraphics[scale=0.8]{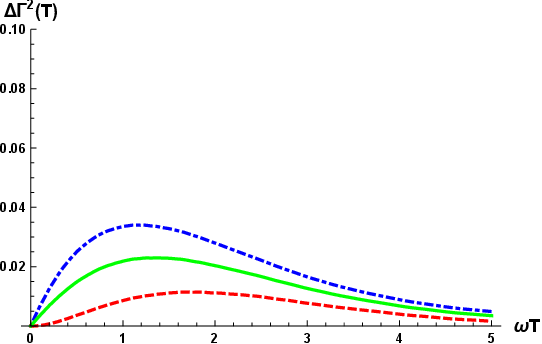}
\caption{$\Delta \Gamma=|\Gamma^2_{\rm ex}(T)-\Gamma^2(T)|$ for the parameters $\omega=1$, $\sigma=1.5$, $\lambda=0.1$. 
The dot-dashed line was computed in the It\^o interpretation $\alpha=0$, the solid line, in the Stratonovich prescription $\alpha=1/2$, 
while the dashed line was computed in the H\"anggi-Klimontovich prescription $\alpha=1$.}
\label{fig:Error}
\end{figure}

We quantified the observed deviation by computing the di\-fference 
$\Delta\Gamma(T)=|\Gamma_{\rm ex}^2(T)-\Gamma^2(T)|$  for  different stochastic prescriptions. In Figure~\ref{fig:Error}, we depict 
$\Delta\Gamma$ for   $\sigma=1.5$ and  $\lambda=0.1$.  The dot-dashed curve was computed in the It\^o prescription ($\alpha=0$), 
the solid line was evaluated in the Stratonovich prescription ($\alpha=1/2$) and  the
 anti-It\^o interpretation ($\alpha=1$) is shown by the  dashed line. 
Clearly, although the errors are very small for these values of the parameters,  we get our best result in the anti-It\^o prescription. 
The reason is that in this prescription, the exact asymptotic probability density is Gaussian, the same one of our approximation. 
The exact asymptotic expression for any other value of $\alpha$ is, in general, not Gaussian. This fact is reflected in Figure~\ref{fig:Error}, 
in the residual values of the curves for $\omega T>>1$.

\section{Discussion and conclusions}
\label{sec:discussion}
In this paper, we have analyzed the weak noise approximation for computing conditional probabilities in a multiplicative stochastic process. 
We noted that this problem is equivalent to compute the propagator of  a quantum problem which describes a particle with variable mass in an external 
potential. Indeed, the diffusion function $g(x)$, in the classical stochastic model,  plays the role of the inverse mass  in the quantum 
equivalent problem ($m(x)=1/g^2(x)$). On the other hand, the necessary stochastic prescriptions used to properly define the Wiener integral have 
their counterpart in the ordering problem in quantum mechanics. The position-dependent mass mixes non-commuting operators ($x$ and $p$) in the 
kinetic term of the Hamiltonian. 
In the stochastic process, the Stratonovich prescription is equivalent to the Weyl order for the quantum associated problem. 
In this paper, we have presented the computation of the propagator using  a "generalized Stratonovich prescription", parametrized by a 
continuous parameter $0\leq\alpha\leq 1$, which contains the most popular stochastic interpretations as particular cases.

The direct implementation of the usual weak noise, or semi-classical, approximation for this type of systems is quite involved. 
In fact, when performing the saddle-point approximation plus quadratic fluctuations, we see that fluctuations are not Gaussian, due 
to the multiplicative noise effects on the functional integration measure. On the other hand, the fluctuation operator is very elaborate  
because of the time dependence of the diffusion function (or equivalently, the mass). 
To solve this problem and to effectively integrate fluctuations, we have performed a  time reparametrization weighted with the diffusion 
function, computed at the classical solution of the equation of motion, $g(x_{cl}(t))$.  The net effect  is to transform the problem of 
computing fluctuations of a multiplicative process into an equivalent additive noise problem.  This is  one of the main contributions of 
this paper. 

The transformation of a multiplicative noise process into an equivalent additive one  can be done, at least for a single variable, 
by means of  a nonlinear transformation of the stochastic variable. This fact  is very well known and gives rise to all the subtleties of 
different types of stochastic calculus.  In higher dimensions, this transformation can be done only in very particular cases and it 
is quite problematic in the context of the path integral representation, even in one dimension~\cite{Lecomte-Leticia2017}.
The transformation we have presented in this paper is of a different type, since it is not a variable transformation, but a \emph{local time 
reparametrization}. Interestingly enough, a similar, however different, type of  time reparametrization  was previously used in the context 
of path integral to  relate the harmonic oscillator propagator with  the free particle propagator~\cite{Luca2004}. 

In some sense, what we are doing is  rectifying the multiplicative character of the noise by locally changing the way we measure time. 
A   similar related idea was recently proposed in Ref.~\onlinecite{Rubin2014}. In that reference, a time reparametrization equivalent to 
Eq.~(\ref{eq:reparametrization}) was proposed, but keeping the stochastic character of the variable $x$. In this way,  a new ``stochastic time"  
was introduced, allowing to change the multiplicative process into an additive one. Differently, in our approach, the time transformation,  
Eq.~(\ref{eq:reparametrization}), is not stochastic  since the diffusion function is evaluated at the solution of the saddle-point equation.  
The classical solution provides a well-defined protocol to rectify the multiplicative noise. Of course, this procedure  works for computing
quadratic fluctuations. Probably, it would be necessary to correct it, order by order, if we pretend to compute higher order fluctuations. An additional  advantage of the time reparametrization technique  is  that it can be straightforwardly implemented in  higher dimensional systems. 

We illustrated the procedure by showing the explicit computation of the propagator of the simplest nontrivial model. We have considered
an overdamped harmonic oscillator with nonlinear multiplicative noise. We have solved the saddle-point equation in 
the weak multiplicative noise regime and we computed fluctuations using the proposed time reparametrization. We showed how the 
multiplicative noise affects the conditional probability and we have shown short and long times limits in order to check the results. 
We have also compared the analytic approximate propagator with the  numerical solution of the Fokker-Planck equation finding an excellent match, 
even well beyond the  parameters region where the approximation is expected to work.

Having this powerful technique in hands, it is now possible to attack more involved problems, such as a potential with several minima, 
in order to address problems like stochastic resonance in systems with multiplicative noise. We hope to report on this subject in the near future. 

\appendix
\section{Saddle-point approximation}
\label{Ap:SP}
The first step in the weak noise approximation  is to solve the saddle-point equation, Eq.~(\ref{eq:SadlePoint}) or, 
equivalently, Eq.~(\ref{eq:firstIntegral}).
By using the expression for the equilibrium potential $U_{eq}(x)$ given by Eq.~(\ref{eq:UeqOs}), we can build  $V(x)$ from Eq.~(\ref{eq:V}). 
It can be approximated as 
\begin{equation}
V(x)=-\frac{\omega\sigma^2}{4}+\left[ \frac{\omega^2}{8}-\frac{\lambda^2 \omega}{2}\sigma^2(1+2\alpha)\right] x^2
\label{eqa:V}
\end{equation}
 where, in the spirit of the weak noise approximation, we have kept terms up to order $\omega\sigma^2$. We have also considered 
 a weak multiplicative effect, $|\lambda x|<<1$. That is, we compute the potential in a distance range  $x^2<< 1/\lambda^2$, where   
 the diffusion function $g(x)$ is not very far away from its additive value $g=1$.
 In the same approximation, we compute $V_{\rm eff}(x)$  from Eq.~(\ref{eq:Veff}), obtaining
 \begin{equation}
 V_{\rm eff}(x)=H-\frac{\omega\sigma^2}{4}+\frac{\Omega^2}{2} x^2 \; ,
\label{eqa:VeffOsc}
\end{equation}
 with 
 \begin{equation}
 \Omega^2=\frac{\omega^2}{4}+\lambda^2\left(4 H-2\omega\sigma^2(1+\alpha)\right) \; .
 \label{eqa:Omega}
 \end{equation}
 With this expression for $V_{\rm eff}$,  we solve Eq.~(\ref{eq:firstIntegral}),  with the boundary conditions $x_{cl}(0)=0$ and  $x_{cl}(T)=x$. 
 We immediately find the classical solution  
 \begin{equation}
 x_{cl}(t)=\frac{x}{\sinh\Omega T} \sinh(\Omega t) \;.
 \label{eqa:xclOsc}
 \end{equation}
 In order to have a completely defined solution, we need to determine the  constant $H$, contained in the frequency $\Omega$.
To determine it, we observe from Eq.~(\ref{eq:vsquare}) that  $\dot x_{cl}(0)=\sqrt{2 H-\omega\sigma^2/2}$. Using the solution 
Eq.~(\ref{eqa:xclOsc}) we obtain the transcendental equation 
 \begin{equation}
 H=\frac{\Omega^2(H) x^2}{2\sinh^2[\Omega(H)T]}+\frac{\omega\sigma^2}{4}\; .
\label{eqa:HOmega} 
 \end{equation}
The solution of equations~(\ref{eqa:Omega}) and~(\ref{eqa:HOmega}) for $H$ and $\Omega$ completely determines the classical solutions
$x_{cl}$, Eq.~(\ref{eqa:xclOsc}).
Although this equation can only be solved numerically, we observe that, within the range of our approximation, an  analytic expression can be found. 
Using a perturbative recurrent procedure, we find, to  order  $\lambda^2 x^2$,  
 \begin{align}
 H=&\frac{\omega\sigma^2}{4}+\frac{\omega^2 x^2}{8\sinh^2(\omega T/2)} 
\label{eqa:H2}
\\
 &-\frac{\lambda^2x^2\omega\sigma^2(1+2\alpha) }{2\sinh^2(\omega T/2)}\left(1-\frac{\omega T}{2}\coth\frac{\omega T}{2}\right)+
  O(\lambda^4 x^4) \;
  \nonumber \\
  \Omega = & \frac{\omega T}{2}-\lambda^2\left(\sigma^2(1+2\alpha)-\frac{\omega x^2}{2\sinh^2(\omega T/2)}\right)+O(\lambda^4 x^4)
  \label{eqa:Omega2}
 \end{align}
 
Now, we are in condition to compute $S_{cl}$ from  Eq.~(\ref{eq:Scl}). Using  Eq.~(\ref{eq:vsquare}) we find the simpler equation
\begin{equation}
S_{cl}=2\int_{0}^T V(x_{cl})\; dt+HT \; .
\label{eqa:Sclsimp}
\end{equation}
Replacing $x_{cl}(t)$ into Eq.~(\ref{eqa:Sclsimp}), performing the time integral and using the  expressions of Eqs.~(\ref{eqa:H2}) 
and~(\ref{eqa:Omega2}),   we find to quadratic order in $\lambda x$, 
\begin{align}
 S_{cl}=&\frac{\omega x^2}{4} \coth(\omega T/2)-\sigma^2\frac{\omega T}{4}
 \label{eqa:Scl} \\
 &-\frac{\sigma^2 \lambda^2 x^2(1+2\alpha)}{2}\left(1-\frac{\omega T}{\sinh(\omega T)}\right) \coth(\omega T/2)
 \nonumber
 \end{align} 
The first line of Eq.~(\ref{eqa:Scl})  is the usual result for the classical action of the harmonic oscillator. 
The second line, proportional to $\lambda^2x^2 $, codifies information about the  multiplicative noise.

\section{Fluctuations}
\label{Ap:fluctuations}
To compute fluctuations, we need to evaluate the prefactor in Eq.~(\ref{eq:PropagatorWeaknoise}),  ${\cal N}\equiv\det^{-1/2}{\Sigma(0,\tau_f)}$,
 where $\Sigma(0,\tau_f)$ is the time reparametrized fluctuation operator given by Eq.~(\ref{eq:Sigma}) with the fluctuation potential, 
\begin{equation}
 W(x_{cl})=\frac{\omega^2}{4}+\frac{\lambda^2 x^2\omega^2}{2\sinh^2(\omega T/2)} \left(1+8 \sinh^2(\omega \tau/2)\right)  .
 \label{eqa:Wosc1}
 \end{equation} 
 To obtain Eq.~(\ref{eqa:Wosc1}) we have discarded any term proportional to the noise intensity $\sigma^2$. 
 This is so because, at this level of approximation, the prefactor  will contribute, upon exponentiation,  to order $\sigma^2$ to the 
 effective action. Then, any term proportional to $\sigma^2$ in the prefactor will contribute to the order $\sigma^4$ in the effective action.
 
To compute the prefactor we multiply and divide by the determinant of the free particle operator $\partial_\tau^2$. Then, we need to compute
\begin{equation}
{\cal N}^{-1}(\tau_f)=\sqrt{\det \partial_\tau^2}\; \sqrt{\frac{\det{\Sigma(0,\tau_f)}}{\det \partial_\tau^2}} \,.
\label{eqa:prefactor}
\end{equation} 
We recall that the variable $\tau$ is the reparametrized time given by Eq.~(\ref{eq:reparametrization}).  
In our particular example, the final reparametrized time $\tau_f$ is written in terms of $T$ as  
\begin{equation}
\tau_f=T\left\{1+\frac{\lambda^2 x^2}{\sinh^2(\omega T/2)}\left(\frac{\sinh\omega T}{\omega T}-1\right) \right\} .
\label{eqa:tauf}
\end{equation}
 
The second factor of Eq.~(\ref{eqa:prefactor}) can be computed by means of the Gelfand-Yaglom theorem~\cite{Dunne2008}. That is, 
\begin{equation}
\frac{\det\Sigma(0,\tau)}{\det \partial_\tau^2}=\frac{\psi(\tau_f)}{\tau_f} \ ,
\label{eqa:detSigmaPrime}
\end{equation} 
where $\psi(\tau)$ is the solution of the homogeneous equation
 \begin{equation}
 -\frac{d^2\psi(\tau)}{d\tau^2}+W(x_{cl})\psi(\tau)=0 \ ,
\label{eqa:Diff-Fluctuations}
 \end{equation}
 with the initial conditions $\psi(0)=0$, $d\psi(0)/d\tau=1$.
Thus, the evaluation of  the ratio of determinants  is reduced to solve an ordinary homogeneous second order differential equation, 
with initial conditions. 
 Due to the extremely nonlinear character of the fluctuation potential $W(x_{cl})$, this equation cannot be solved using elementary functions. 
 Fortunately, the solution of Eq.~(\ref{eqa:Diff-Fluctuations}) can be expressed in a closed form in terms of Mathieu 
 functions~\cite{abramowitz-1964,gradshteyn2007}.  
 The result is 
 \begin{equation}
 \psi(\tau)= -\frac{2 i}{\omega}\frac{Se_r(i\omega\tau/2,q)}{Se_r'(0,q)}
\label{eqa:psi} 
 \end{equation} 
 where $Se_r(z,q)$ is the odd Mathieu function of imaginary argument   $z=i\omega \tau$. $Se_r'(z,q)$ is the first derivative of the Mathieu function with respect to $z$. The parameters $r$ and $q$ are given by 
\begin{align} 
 r&=1-12 ~\frac{\lambda^2 x^2}{2\sinh(\omega T/2)} \; ,  \\
 q&= -~8~\frac{\lambda^2 x^2}{2\sinh(\omega T/2)} \; .
\end{align}
Interestingly, for additive noise ($\lambda=0$), the parameters are $r=1$ and $q=0$. For these values of the parameters, 
the Mathieu functions reduce to elementary functions, 
\begin{equation}
Se_1(i\frac{\omega\tau}{2},0)=i\sinh \frac{\omega\tau}{2}\; .
\end{equation}
Then, we can expand Eq.~(\ref{eqa:psi}) in powers of 
 $\lambda^2x^2$.  To explicitly perform this expansion we only need two properties of the Mathieu functions~\cite{abramowitz-1964,gradshteyn2007}. 
 The first of these properties is 
 \begin{equation}
 Se_1(q,z)\sim \sin z-\frac{1}{8}\sin(3 z) q + O(q^2) \;,
 \end{equation} 
 and the second one, 
\begin{equation}
Se_r(0,z)=\frac{\sin(\sqrt{r} z)}{\sqrt{r}} \;.
\end{equation} 
Using these expressions we find, for $\psi(\tau)$,
 \begin{align}
 \psi(\tau) = & ~\frac{2}{\omega}\sinh(\omega \tau/2) \nonumber \\
 &+\frac{6}{\omega}\frac{\lambda^2 x^2}{\sinh(\omega T/2)}
 \left\{\frac{1}{2}\sinh(\omega\tau/2)+\frac{1}{6}\sinh(3\omega\tau/2)\right. \nonumber \\
 &\left. -~\frac{\omega \tau}{2}\cosh(\omega\tau/2)\right\} ~ + ~O(\lambda^4 x^4) \; .
 \end{align}
Replacing this result into Eq.~(\ref{eqa:detSigmaPrime}), we find the determinant ratio that should be evaluated at the final time $\tau_f$. 

To complete the calculation, we need to go back to ordinary time using Eq.~(\ref{eqa:tauf}). Keeping the leading-order term in $\lambda^2x^2$ we find
\begin{widetext}
\begin{align}
 \sqrt{\frac{\det \partial_\tau^2}{\det\Sigma(0,\tau)}} = &~\sqrt{\frac{\omega T}{2\sinh(\omega T/2)}}\left\{ 1- \frac{\lambda^2x^2}{2\sinh^2(\omega T/2)}  \right. \nonumber \\
 &\times \left.\left[\frac{3}{2}+\frac{1}{2}\frac{\sinh(3\omega T/2)}{\sinh(\omega T/2)}-\frac{3}{2} \omega T \mbox{coth}(\omega T/2)
-\left(\frac{\sinh(\omega T)}{\omega T}-1\right)\left(1-\frac{\omega T}{2} \mbox{coth}(\omega T/2)\right)
  \right]\right\} \; .
  \label{eqa:detsResult}
 \end{align}
\end{widetext}

Now, to finished the calculation of  ${\cal N}(T)$ in Eq.~(\ref{eqa:prefactor}), it is necessary to carefully evaluate $\det\partial^2_\tau$. 
Usually, this factor is absorbed in a global normalization constant. However, in our case, due to the reparametrization  of time , 
this procedure should be done with great care. Computing the determinant thorough the eigenvalues of the operator $\partial^2_\tau$, 
it is simple to show that
\begin{equation}
\det\partial^2_\tau=\prod_n \frac{\pi^2 n^2}{\tau^2_f} \, .
\end{equation}
Going back to the original time variable, using Eq.~(\ref{eqa:tauf}) we find, 
\begin{align}
\det\partial^2_\tau=&\left(\prod_n \frac{\pi^2 n^2}{T^2}\right)\nonumber \\
 &\times \prod_n \frac{1}{\left[1+\frac{\lambda^2 x^2}{\sinh^2(\omega T/2)}\left(\frac{\sinh\omega T}{\omega T}-1\right)\right]^2}
\end{align}
The first factor can indeed be absorbed in a global normalization constant ${\cal N}'$. However, the second factor depends on $x$ and  cannot be 
ignored. Using the zeta-function regularization~\cite{zfunction1994} we find, 
 \begin{equation}
\det\partial^2_\tau={\cal N}'\left[ 1+\frac{\lambda^2 x^2}{\sinh^2(\omega T/2)}\left(\frac{\sinh\omega T}{\omega T}-1\right)\right]
\end{equation}
Replacing this expression in Eq.~(\ref{eqa:prefactor}), using Eq.~(\ref{eqa:detsResult}) and re-expanding to leading order in $\lambda^2 x^2$, 
we finally find 
\begin{align}
 &{\cal N}(T)={\cal N}'\sqrt{\frac{\omega T}{2\sinh(\omega T/2)}}\left\{ 1- \frac{\lambda^2x^2}{2\sinh^2(\omega T/2)} \right.
 \label{eqa:prefactorResult} \\
  & \left.
 \left[\frac{3}{2}+\frac{1}{2}\frac{\sinh(3\omega T/2)}{\sinh(\omega T/2)}+ \frac{\omega T}{2}  \mbox{coth}(\omega T/2)\left(\frac{\sinh(\omega T)}{\omega T}-4\right)
  \right]\right\} \nonumber \\
  \nonumber
   \end{align}
which coincides with Eq.~(\ref{eq:detsResult}).
 
\acknowledgments
The Brazilian agencies, {\em Funda\c c\~ao de Amparo \`a Pesquisa do Rio
de Janeiro} (FAPERJ), {\em Conselho Nacional de Desenvolvimento Cient\'\i
fico e Tecnol\'ogico} (CNPq) and {\em Coordena\c c\~ao  de Aperfei\c coamento de Pessoal de N\'\i vel Superior}  (CAPES) - Finance Code 001,  are acknowledged  for partial financial support.
DGB wants to acknowledge the Physics Institute of La Plata, Argentina (IFLP) and  the Physics Department 
``Juan José Giambiagi'', of the University of Buenos Aires (UBA), Argentina,  for kind hospitality during the preparation of this work. 


\begin{thebibliography}{55}%
\makeatletter
\providecommand \@ifxundefined [1]{%
 \@ifx{#1\undefined}
}%
\providecommand \@ifnum [1]{%
 \ifnum #1\expandafter \@firstoftwo
 \else \expandafter \@secondoftwo
 \fi
}%
\providecommand \@ifx [1]{%
 \ifx #1\expandafter \@firstoftwo
 \else \expandafter \@secondoftwo
 \fi
}%
\providecommand \natexlab [1]{#1}%
\providecommand \enquote  [1]{``#1''}%
\providecommand \bibnamefont  [1]{#1}%
\providecommand \bibfnamefont [1]{#1}%
\providecommand \citenamefont [1]{#1}%
\providecommand \href@noop [0]{\@secondoftwo}%
\providecommand \href [0]{\begingroup \@sanitize@url \@href}%
\providecommand \@href[1]{\@@startlink{#1}\@@href}%
\providecommand \@@href[1]{\endgroup#1\@@endlink}%
\providecommand \@sanitize@url [0]{\catcode `\\12\catcode `\$12\catcode
  `\&12\catcode `\#12\catcode `\^12\catcode `\_12\catcode `\%12\relax}%
\providecommand \@@startlink[1]{}%
\providecommand \@@endlink[0]{}%
\providecommand \url  [0]{\begingroup\@sanitize@url \@url }%
\providecommand \@url [1]{\endgroup\@href {#1}{\urlprefix }}%
\providecommand \urlprefix  [0]{URL }%
\providecommand \Eprint [0]{\href }%
\providecommand \doibase [0]{http://dx.doi.org/}%
\providecommand \selectlanguage [0]{\@gobble}%
\providecommand \bibinfo  [0]{\@secondoftwo}%
\providecommand \bibfield  [0]{\@secondoftwo}%
\providecommand \translation [1]{[#1]}%
\providecommand \BibitemOpen [0]{}%
\providecommand \bibitemStop [0]{}%
\providecommand \bibitemNoStop [0]{.\EOS\space}%
\providecommand \EOS [0]{\spacefactor3000\relax}%
\providecommand \BibitemShut  [1]{\csname bibitem#1\endcsname}%
\let\auto@bib@innerbib\@empty
\bibitem [{\citenamefont {Gardiner}(1996)}]{gardiner}%
  \BibitemOpen
  \bibfield  {author} {\bibinfo {author} {\bibfnamefont {C.~W.}\ \bibnamefont
  {Gardiner}},\ }\href@noop {} {\emph {\bibinfo {title} {Handbook of stochastic
  methods for physics, chemistry and the natural sciences}}}\ (\bibinfo
  {publisher} {Springer-Verlag},\ \bibinfo {address} {Berlin Heidelberg},\
  \bibinfo {year} {1996})\BibitemShut {NoStop}%
\bibitem [{\citenamefont {van Kampen}(2007)}]{vanKampen}%
  \BibitemOpen
  \bibfield  {author} {\bibinfo {author} {\bibfnamefont {N.~G.}\ \bibnamefont
  {van Kampen}},\ }\href@noop {} {\emph {\bibinfo {title} {{Stochastic
  Processes in Physics and Chemistry}}}}\ (\bibinfo  {publisher} {Elsevier},\
  \bibinfo {address} {London, UK},\ \bibinfo {year} {2007})\BibitemShut
  {NoStop}%
\bibitem [{\citenamefont {Freund}\ and\ \citenamefont
  {P{\"o}schel}(2000)}]{Poschel}%
  \BibitemOpen
  \bibfield  {author} {\bibinfo {author} {\bibfnamefont {J.~A.}\ \bibnamefont
  {Freund}}\ and\ \bibinfo {author} {\bibfnamefont {T.}~\bibnamefont
  {P{\"o}schel}},\ }\href@noop {} {\emph {\bibinfo {title} {{Stochastic
  Processes in Physics, Chemistry and Biology}}}}\ (\bibinfo  {publisher}
  {Springer-Verlag},\ \bibinfo {address} {Berlin, Heidelberg},\ \bibinfo {year}
  {2000})\BibitemShut {NoStop}%
\bibitem [{\citenamefont {Murray}(2002)}]{Murray}%
  \BibitemOpen
  \bibfield  {author} {\bibinfo {author} {\bibfnamefont {J.~D.}\ \bibnamefont
  {Murray}},\ }\href@noop {} {\emph {\bibinfo {title} {{Mathematical Biology.
  I. An introduction}}}}\ (\bibinfo  {publisher} {Springer-Verlag},\ \bibinfo
  {address} {Berlin, Heidelberg},\ \bibinfo {year} {2002})\BibitemShut
  {NoStop}%
\bibitem [{\citenamefont {Mantegna}\ and\ \citenamefont
  {Stanley}(2000)}]{Mantegna}%
  \BibitemOpen
  \bibfield  {author} {\bibinfo {author} {\bibfnamefont {R.~N.}\ \bibnamefont
  {Mantegna}}\ and\ \bibinfo {author} {\bibfnamefont {H.~E.}\ \bibnamefont
  {Stanley}},\ }\href@noop {} {\emph {\bibinfo {title} {An introduction to
  econophysics: correlations and complexity in finance}}}\ (\bibinfo
  {publisher} {Cambridge University Press},\ \bibinfo {address} {Cambridge,
  UK},\ \bibinfo {year} {2000})\BibitemShut {NoStop}%
\bibitem [{\citenamefont {Bouchaud}\ and\ \citenamefont
  {Potters}(2003)}]{Bouchaud}%
  \BibitemOpen
  \bibfield  {author} {\bibinfo {author} {\bibfnamefont {J.~P.}\ \bibnamefont
  {Bouchaud}}\ and\ \bibinfo {author} {\bibfnamefont {M.}~\bibnamefont
  {Potters}},\ }\href@noop {} {\emph {\bibinfo {title} {Theory of financial
  risk and derivative pricing: from statistical physics to risk management}}}\
  (\bibinfo  {publisher} {Cambridge University Press},\ \bibinfo {year}
  {2003})\BibitemShut {NoStop}%
\bibitem [{\citenamefont {Crooks}(1999)}]{crooks1999}%
  \BibitemOpen
  \bibfield  {author} {\bibinfo {author} {\bibfnamefont {G.~E.}\ \bibnamefont
  {Crooks}},\ }\href {\doibase 10.1103/PhysRevE.60.2721} {\bibfield  {journal}
  {\bibinfo  {journal} {Phys. Rev. E}\ }\textbf {\bibinfo {volume} {60}},\
  \bibinfo {pages} {2721} (\bibinfo {year} {1999})}\BibitemShut {NoStop}%
\bibitem [{\citenamefont {Zwanzig}(2001)}]{ZwanzigBook2001}%
  \BibitemOpen
  \bibfield  {author} {\bibinfo {author} {\bibfnamefont {R.}~\bibnamefont
  {Zwanzig}},\ }\href@noop {} {\emph {\bibinfo {title} {Nonequilibrium
  statistical mechanics}}}\ (\bibinfo  {publisher} {Oxford University Press,
  USA},\ \bibinfo {year} {2001})\BibitemShut {NoStop}%
\bibitem [{\citenamefont {Seifert}(2008)}]{seifert2008}%
  \BibitemOpen
  \bibfield  {author} {\bibinfo {author} {\bibfnamefont {U.}~\bibnamefont
  {Seifert}},\ }\href {http://dx.doi.org/10.1140/epjb/e2008-00001-9} {\bibfield
   {journal} {\bibinfo  {journal} {The European Physical Journal B - Condensed
  Matter and Complex Systems}\ }\textbf {\bibinfo {volume} {64}},\ \bibinfo
  {pages} {423} (\bibinfo {year} {2008})},\ \bibinfo {note}
  {10.1140/epjb/e2008-00001-9}\BibitemShut {NoStop}%
\bibitem [{\citenamefont {Lan\c{c}on}\ \emph {et~al.}(2001)\citenamefont
  {Lan\c{c}on}, \citenamefont {Batrouni}, \citenamefont {Lobry},\ and\
  \citenamefont {Ostrowsky}}]{Lancon2001}%
  \BibitemOpen
  \bibfield  {author} {\bibinfo {author} {\bibfnamefont {P.}~\bibnamefont
  {Lan\c{c}on}}, \bibinfo {author} {\bibfnamefont {G.}~\bibnamefont
  {Batrouni}}, \bibinfo {author} {\bibfnamefont {L.}~\bibnamefont {Lobry}}, \
  and\ \bibinfo {author} {\bibfnamefont {N.}~\bibnamefont {Ostrowsky}},\ }\href
  {http://stacks.iop.org/0295-5075/54/i=1/a=028} {\bibfield  {journal}
  {\bibinfo  {journal} {EPL (Europhysics Letters)}\ }\textbf {\bibinfo {volume}
  {54}},\ \bibinfo {pages} {28} (\bibinfo {year} {2001})}\BibitemShut {NoStop}%
\bibitem [{\citenamefont {Lan\c{c}on}\ \emph {et~al.}(2002)\citenamefont
  {Lan\c{c}on}, \citenamefont {Batrouni}, \citenamefont {Lobry},\ and\
  \citenamefont {Ostrowsky}}]{Lancon2002}%
  \BibitemOpen
  \bibfield  {author} {\bibinfo {author} {\bibfnamefont {P.}~\bibnamefont
  {Lan\c{c}on}}, \bibinfo {author} {\bibfnamefont {G.}~\bibnamefont
  {Batrouni}}, \bibinfo {author} {\bibfnamefont {L.}~\bibnamefont {Lobry}}, \
  and\ \bibinfo {author} {\bibfnamefont {N.}~\bibnamefont {Ostrowsky}},\ }\href
  {http://www.sciencedirect.com/science/article/pii/S0378437101005106}
  {\bibfield  {journal} {\bibinfo  {journal} {Physica A: Statistical Mechanics
  and its Applications}\ }\textbf {\bibinfo {volume} {304}},\ \bibinfo {pages}
  {65 } (\bibinfo {year} {2002})}\BibitemShut {NoStop}%
\bibitem [{\citenamefont {Lau}\ and\ \citenamefont
  {Lubensky}(2007)}]{Lubensky2007}%
  \BibitemOpen
  \bibfield  {author} {\bibinfo {author} {\bibfnamefont {A.~W.~C.}\
  \bibnamefont {Lau}}\ and\ \bibinfo {author} {\bibfnamefont {T.~C.}\
  \bibnamefont {Lubensky}},\ }\href {\doibase 10.1103/PhysRevE.76.011123}
  {\bibfield  {journal} {\bibinfo  {journal} {Phys. Rev. E}\ }\textbf {\bibinfo
  {volume} {76}},\ \bibinfo {pages} {011123} (\bibinfo {year}
  {2007})}\BibitemShut {NoStop}%
\bibitem [{\citenamefont {Volpe}\ \emph {et~al.}(2010)\citenamefont {Volpe},
  \citenamefont {Helden}, \citenamefont {Brettschneider}, \citenamefont
  {Wehr},\ and\ \citenamefont {Bechinger}}]{Volpe2010}%
  \BibitemOpen
  \bibfield  {author} {\bibinfo {author} {\bibfnamefont {G.}~\bibnamefont
  {Volpe}}, \bibinfo {author} {\bibfnamefont {L.}~\bibnamefont {Helden}},
  \bibinfo {author} {\bibfnamefont {T.}~\bibnamefont {Brettschneider}},
  \bibinfo {author} {\bibfnamefont {J.}~\bibnamefont {Wehr}}, \ and\ \bibinfo
  {author} {\bibfnamefont {C.}~\bibnamefont {Bechinger}},\ }\href {\doibase
  10.1103/PhysRevLett.104.170602} {\bibfield  {journal} {\bibinfo  {journal}
  {Phys. Rev. Lett.}\ }\textbf {\bibinfo {volume} {104}},\ \bibinfo {pages}
  {170602} (\bibinfo {year} {2010})}\BibitemShut {NoStop}%
\bibitem [{\citenamefont {Brettschneider}\ \emph {et~al.}(2011)\citenamefont
  {Brettschneider}, \citenamefont {Volpe}, \citenamefont {Helden},
  \citenamefont {Wehr},\ and\ \citenamefont {Bechinger}}]{Volpe2011}%
  \BibitemOpen
  \bibfield  {author} {\bibinfo {author} {\bibfnamefont {T.}~\bibnamefont
  {Brettschneider}}, \bibinfo {author} {\bibfnamefont {G.}~\bibnamefont
  {Volpe}}, \bibinfo {author} {\bibfnamefont {L.}~\bibnamefont {Helden}},
  \bibinfo {author} {\bibfnamefont {J.}~\bibnamefont {Wehr}}, \ and\ \bibinfo
  {author} {\bibfnamefont {C.}~\bibnamefont {Bechinger}},\ }\href {\doibase
  10.1103/PhysRevE.83.041113} {\bibfield  {journal} {\bibinfo  {journal} {Phys.
  Rev. E}\ }\textbf {\bibinfo {volume} {83}},\ \bibinfo {pages} {041113}
  (\bibinfo {year} {2011})}\BibitemShut {NoStop}%
\bibitem [{\citenamefont {Garc\'{\i}a-Palacios}\ and\ \citenamefont
  {L\'azaro}(1998)}]{GarciaPalacios1998}%
  \BibitemOpen
  \bibfield  {author} {\bibinfo {author} {\bibfnamefont {J.~L.}\ \bibnamefont
  {Garc\'{\i}a-Palacios}}\ and\ \bibinfo {author} {\bibfnamefont {F.~J.}\
  \bibnamefont {L\'azaro}},\ }\href {\doibase 10.1103/PhysRevB.58.14937}
  {\bibfield  {journal} {\bibinfo  {journal} {Phys. Rev. B}\ }\textbf {\bibinfo
  {volume} {58}},\ \bibinfo {pages} {14937} (\bibinfo {year}
  {1998})}\BibitemShut {NoStop}%
\bibitem [{\citenamefont {Aron}\ \emph {et~al.}(2014)\citenamefont {Aron},
  \citenamefont {Barci}, \citenamefont {Cugliandolo}, \citenamefont {Arenas},\
  and\ \citenamefont {Lozano}}]{Aron2014}%
  \BibitemOpen
  \bibfield  {author} {\bibinfo {author} {\bibfnamefont {C.}~\bibnamefont
  {Aron}}, \bibinfo {author} {\bibfnamefont {D.~G.}\ \bibnamefont {Barci}},
  \bibinfo {author} {\bibfnamefont {L.~F.}\ \bibnamefont {Cugliandolo}},
  \bibinfo {author} {\bibfnamefont {Z.~G.}\ \bibnamefont {Arenas}}, \ and\
  \bibinfo {author} {\bibfnamefont {G.~S.}\ \bibnamefont {Lozano}},\ }\href
  {http://stacks.iop.org/1742-5468/2014/i=9/a=P09008} {\bibfield  {journal}
  {\bibinfo  {journal} {Journal of Statistical Mechanics: Theory and
  Experiment}\ }\textbf {\bibinfo {volume} {2014}},\ \bibinfo {pages} {P09008}
  (\bibinfo {year} {2014})}\BibitemShut {NoStop}%
\bibitem [{\citenamefont {Arenas}\ \emph {et~al.}(2018)\citenamefont {Arenas},
  \citenamefont {Barci},\ and\ \citenamefont {Moreno}}]{Arenas2018}%
  \BibitemOpen
  \bibfield  {author} {\bibinfo {author} {\bibfnamefont {G.}~\bibnamefont
  {Arenas}}, \bibinfo {author} {\bibfnamefont {D.~G.}\ \bibnamefont {Barci}}, \
  and\ \bibinfo {author} {\bibfnamefont {M.~V.}\ \bibnamefont {Moreno}},\
  }\href {\doibase https://doi.org/10.1016/j.physa.2018.06.126} {\bibfield
  {journal} {\bibinfo  {journal} {Physica A: Statistical Mechanics and its
  Applications}\ }\textbf {\bibinfo {volume} {510}},\ \bibinfo {pages} {98 }
  (\bibinfo {year} {2018})}\BibitemShut {NoStop}%
\bibitem [{\citenamefont {Hinrichsen}(2000)}]{Hinrichsen2000}%
  \BibitemOpen
  \bibfield  {author} {\bibinfo {author} {\bibfnamefont {H.}~\bibnamefont
  {Hinrichsen}},\ }\href {\doibase 10.1080/00018730050198152} {\bibfield
  {journal} {\bibinfo  {journal} {Advances in Physics}\ }\textbf {\bibinfo
  {volume} {49}},\ \bibinfo {pages} {815} (\bibinfo {year} {2000})},\ \Eprint
  {http://arxiv.org/abs/https://doi.org/10.1080/00018730050198152}
  {https://doi.org/10.1080/00018730050198152} \BibitemShut {NoStop}%
\bibitem [{\citenamefont {Van~den Broeck}\ \emph {et~al.}(1994)\citenamefont
  {Van~den Broeck}, \citenamefont {Parrondo},\ and\ \citenamefont
  {Toral}}]{Parrondo1994}%
  \BibitemOpen
  \bibfield  {author} {\bibinfo {author} {\bibfnamefont {C.}~\bibnamefont
  {Van~den Broeck}}, \bibinfo {author} {\bibfnamefont {J.~M.~R.}\ \bibnamefont
  {Parrondo}}, \ and\ \bibinfo {author} {\bibfnamefont {R.}~\bibnamefont
  {Toral}},\ }\href {\doibase 10.1103/PhysRevLett.73.3395} {\bibfield
  {journal} {\bibinfo  {journal} {Phys. Rev. Lett.}\ }\textbf {\bibinfo
  {volume} {73}},\ \bibinfo {pages} {3395} (\bibinfo {year}
  {1994})}\BibitemShut {NoStop}%
\bibitem [{\citenamefont {Castro}\ \emph {et~al.}(1995)\citenamefont {Castro},
  \citenamefont {S\'anchez},\ and\ \citenamefont {Wio}}]{CastroWio1995}%
  \BibitemOpen
  \bibfield  {author} {\bibinfo {author} {\bibfnamefont {F.}~\bibnamefont
  {Castro}}, \bibinfo {author} {\bibfnamefont {A.~D.}\ \bibnamefont
  {S\'anchez}}, \ and\ \bibinfo {author} {\bibfnamefont {H.~S.}\ \bibnamefont
  {Wio}},\ }\href {\doibase 10.1103/PhysRevLett.75.1691} {\bibfield  {journal}
  {\bibinfo  {journal} {Phys. Rev. Lett.}\ }\textbf {\bibinfo {volume} {75}},\
  \bibinfo {pages} {1691} (\bibinfo {year} {1995})}\BibitemShut {NoStop}%
\bibitem [{\citenamefont {Carrillo}\ \emph {et~al.}(2003)\citenamefont
  {Carrillo}, \citenamefont {Iba\~nes}, \citenamefont {Garc\'{\i}a-Ojalvo},
  \citenamefont {Casademunt},\ and\ \citenamefont {Sancho}}]{Sancho2003}%
  \BibitemOpen
  \bibfield  {author} {\bibinfo {author} {\bibfnamefont {O.}~\bibnamefont
  {Carrillo}}, \bibinfo {author} {\bibfnamefont {M.}~\bibnamefont {Iba\~nes}},
  \bibinfo {author} {\bibfnamefont {J.}~\bibnamefont {Garc\'{\i}a-Ojalvo}},
  \bibinfo {author} {\bibfnamefont {J.}~\bibnamefont {Casademunt}}, \ and\
  \bibinfo {author} {\bibfnamefont {J.~M.}\ \bibnamefont {Sancho}},\ }\href
  {\doibase 10.1103/PhysRevE.67.046110} {\bibfield  {journal} {\bibinfo
  {journal} {Phys. Rev. E}\ }\textbf {\bibinfo {volume} {67}},\ \bibinfo
  {pages} {046110} (\bibinfo {year} {2003})}\BibitemShut {NoStop}%
\bibitem [{\citenamefont {Jafarpour}\ \emph {et~al.}(2015)\citenamefont
  {Jafarpour}, \citenamefont {Biancalani},\ and\ \citenamefont
  {Goldenfeld}}]{Goldenfeld2015}%
  \BibitemOpen
  \bibfield  {author} {\bibinfo {author} {\bibfnamefont {F.}~\bibnamefont
  {Jafarpour}}, \bibinfo {author} {\bibfnamefont {T.}~\bibnamefont
  {Biancalani}}, \ and\ \bibinfo {author} {\bibfnamefont {N.}~\bibnamefont
  {Goldenfeld}},\ }\href {\doibase 10.1103/PhysRevLett.115.158101} {\bibfield
  {journal} {\bibinfo  {journal} {Phys. Rev. Lett.}\ }\textbf {\bibinfo
  {volume} {115}},\ \bibinfo {pages} {158101} (\bibinfo {year}
  {2015})}\BibitemShut {NoStop}%
\bibitem [{\citenamefont {Barci}\ \emph {et~al.}(2016)\citenamefont {Barci},
  \citenamefont {Arenas},\ and\ \citenamefont
  {Moreno}}]{BarciMiguelZochil2016}%
  \BibitemOpen
  \bibfield  {author} {\bibinfo {author} {\bibfnamefont {D.~G.}\ \bibnamefont
  {Barci}}, \bibinfo {author} {\bibfnamefont {Z.~G.}\ \bibnamefont {Arenas}}, \
  and\ \bibinfo {author} {\bibfnamefont {M.~V.}\ \bibnamefont {Moreno}},\
  }\href {http://stacks.iop.org/0295-5075/113/i=1/a=10009} {\bibfield
  {journal} {\bibinfo  {journal} {EPL (Europhysics Letters)}\ }\textbf
  {\bibinfo {volume} {113}},\ \bibinfo {pages} {10009} (\bibinfo {year}
  {2016})}\BibitemShut {NoStop}%
\bibitem [{\citenamefont {Benzi}\ \emph {et~al.}(1981)\citenamefont {Benzi},
  \citenamefont {Sutera},\ and\ \citenamefont {Vulpiani}}]{Benzi1981}%
  \BibitemOpen
  \bibfield  {author} {\bibinfo {author} {\bibfnamefont {R.}~\bibnamefont
  {Benzi}}, \bibinfo {author} {\bibfnamefont {A.}~\bibnamefont {Sutera}}, \
  and\ \bibinfo {author} {\bibfnamefont {A.}~\bibnamefont {Vulpiani}},\ }\href
  {http://stacks.iop.org/0305-4470/14/i=11/a=006} {\bibfield  {journal}
  {\bibinfo  {journal} {Journal of Physics A: Mathematical and General}\
  }\textbf {\bibinfo {volume} {14}},\ \bibinfo {pages} {L453} (\bibinfo {year}
  {1981})}\BibitemShut {NoStop}%
\bibitem [{\citenamefont {Benzi}\ \emph {et~al.}(1983)\citenamefont {Benzi},
  \citenamefont {Parisi}, \citenamefont {Sutera},\ and\ \citenamefont
  {Vulpiani}}]{Parisi1983}%
  \BibitemOpen
  \bibfield  {author} {\bibinfo {author} {\bibfnamefont {R.}~\bibnamefont
  {Benzi}}, \bibinfo {author} {\bibfnamefont {G.}~\bibnamefont {Parisi}},
  \bibinfo {author} {\bibfnamefont {A.}~\bibnamefont {Sutera}}, \ and\ \bibinfo
  {author} {\bibfnamefont {A.}~\bibnamefont {Vulpiani}},\ }\href {\doibase
  10.1137/0143037} {\bibfield  {journal} {\bibinfo  {journal} {SIAM Journal on
  Applied Mathematics}\ }\textbf {\bibinfo {volume} {43}},\ \bibinfo {pages}
  {565} (\bibinfo {year} {1983})},\ \Eprint
  {http://arxiv.org/abs/https://doi.org/10.1137/0143037}
  {https://doi.org/10.1137/0143037} \BibitemShut {NoStop}%
\bibitem [{\citenamefont {{Wio, H. S.}}\ and\ \citenamefont {{Deza, R.
  R.}}(2007)}]{Wio2007}%
  \BibitemOpen
  \bibfield  {author} {\bibinfo {author} {\bibnamefont {{Wio, H. S.}}}\ and\
  \bibinfo {author} {\bibnamefont {{Deza, R. R.}}},\ }\href {\doibase
  10.1140/epjst/e2007-00173-0} {\bibfield  {journal} {\bibinfo  {journal} {Eur.
  Phys. J. Special Topics}\ }\textbf {\bibinfo {volume} {146}},\ \bibinfo
  {pages} {111} (\bibinfo {year} {2007})}\BibitemShut {NoStop}%
\bibitem [{\citenamefont {Wio}\ \emph {et~al.}(2002)\citenamefont {Wio},
  \citenamefont {Bouzat},\ and\ \citenamefont {von Haeften}}]{Wio2002}%
  \BibitemOpen
  \bibfield  {author} {\bibinfo {author} {\bibfnamefont {H.}~\bibnamefont
  {Wio}}, \bibinfo {author} {\bibfnamefont {S.}~\bibnamefont {Bouzat}}, \ and\
  \bibinfo {author} {\bibfnamefont {B.}~\bibnamefont {von Haeften}},\ }\href
  {\doibase http://dx.doi.org/10.1016/S0378-4371(02)00493-4} {\bibfield
  {journal} {\bibinfo  {journal} {Physica A: Statistical Mechanics and its
  Applications}\ }\textbf {\bibinfo {volume} {306}},\ \bibinfo {pages} {140 }
  (\bibinfo {year} {2002})},\ \bibinfo {note} {invited Papers from the 21th
  \{IUPAP\} International Conference on St atistical Physics}\BibitemShut
  {NoStop}%
\bibitem [{\citenamefont {Sivak}\ \emph {et~al.}(2014)\citenamefont {Sivak},
  \citenamefont {Chodera},\ and\ \citenamefont {Crooks}}]{Sivak2014}%
  \BibitemOpen
  \bibfield  {author} {\bibinfo {author} {\bibfnamefont {D.~A.}\ \bibnamefont
  {Sivak}}, \bibinfo {author} {\bibfnamefont {J.~D.}\ \bibnamefont {Chodera}},
  \ and\ \bibinfo {author} {\bibfnamefont {G.~E.}\ \bibnamefont {Crooks}},\
  }\href {\doibase https://doi.org/10.1016/j.bpj.2013.11.2269} {\bibfield
  {journal} {\bibinfo  {journal} {Biophysical Journal}\ }\textbf {\bibinfo
  {volume} {106}},\ \bibinfo {pages} {403a} (\bibinfo {year}
  {2014})}\BibitemShut {NoStop}%
\bibitem [{\citenamefont {Goldenfeld}(1992)}]{Goldenfeld}%
  \BibitemOpen
  \bibfield  {author} {\bibinfo {author} {\bibfnamefont {N.}~\bibnamefont
  {Goldenfeld}},\ }\href@noop {} {\emph {\bibinfo {title} {{Lectures On Phase
  Transitions And The Renormalization Group}}}}\ (\bibinfo  {publisher}
  {Frontiers in Physics v.85, Perseus Books Publishing L.L.C.},\ \bibinfo
  {address} {New York, USA},\ \bibinfo {year} {1992})\BibitemShut {NoStop}%
\bibitem [{\citenamefont {Wio}(2013)}]{WioBook2013}%
  \BibitemOpen
  \bibfield  {author} {\bibinfo {author} {\bibfnamefont {H.}~\bibnamefont
  {Wio}},\ }\href {https://books.google.com.br/books?id= vKStkQEACAAJ} {\emph
  {\bibinfo {title} {Path Integrals for Stochastic Processes: An
  Introduction}}}\ (\bibinfo  {publisher} {World Scientific},\ \bibinfo {year}
  {2013})\BibitemShut {NoStop}%
\bibitem [{\citenamefont {Janssen}(1992)}]{Janssen-RG}%
  \BibitemOpen
  \bibfield  {author} {\bibinfo {author} {\bibfnamefont {H.~K.}\ \bibnamefont
  {Janssen}},\ }\href@noop {} {\emph {\bibinfo {title} {From phase transitions
  to chaos: Topics in Modern Statistical Physics}}}\ (\bibinfo  {publisher}
  {World Scientific},\ \bibinfo {address} {Singapore},\ \bibinfo {year}
  {1992})\BibitemShut {NoStop}%
\bibitem [{\citenamefont {Aron}\ \emph {et~al.}(2010)\citenamefont {Aron},
  \citenamefont {Biroli},\ and\ \citenamefont {Cugliandolo}}]{AronLeticia2010}%
  \BibitemOpen
  \bibfield  {author} {\bibinfo {author} {\bibfnamefont {C.}~\bibnamefont
  {Aron}}, \bibinfo {author} {\bibfnamefont {G.}~\bibnamefont {Biroli}}, \ and\
  \bibinfo {author} {\bibfnamefont {L.~F.}\ \bibnamefont {Cugliandolo}},\
  }\href {http://stacks.iop.org/1742-5468/2010/i=11/a=P11018} {\bibfield
  {journal} {\bibinfo  {journal} {Journal of Statistical Mechanics: Theory and
  Experiment}\ }\textbf {\bibinfo {volume} {2010}},\ \bibinfo {pages} {P11018}
  (\bibinfo {year} {2010})}\BibitemShut {NoStop}%
\bibitem [{\citenamefont {Arenas}\ and\ \citenamefont
  {Barci}(2010)}]{arenas2010}%
  \BibitemOpen
  \bibfield  {author} {\bibinfo {author} {\bibfnamefont {Z.~G.}\ \bibnamefont
  {Arenas}}\ and\ \bibinfo {author} {\bibfnamefont {D.~G.}\ \bibnamefont
  {Barci}},\ }\href {\doibase 10.1103/PhysRevE.81.051113} {\bibfield  {journal}
  {\bibinfo  {journal} {Phys. Rev. E}\ }\textbf {\bibinfo {volume} {81}},\
  \bibinfo {pages} {051113} (\bibinfo {year} {2010})}\BibitemShut {NoStop}%
\bibitem [{\citenamefont {Arenas}\ and\ \citenamefont
  {Barci}(2012{\natexlab{a}})}]{arenas2012}%
  \BibitemOpen
  \bibfield  {author} {\bibinfo {author} {\bibfnamefont {Z.~G.}\ \bibnamefont
  {Arenas}}\ and\ \bibinfo {author} {\bibfnamefont {D.~G.}\ \bibnamefont
  {Barci}},\ }\href {\doibase 10.1103/PhysRevE.85.041122} {\bibfield  {journal}
  {\bibinfo  {journal} {Phys. Rev. E}\ }\textbf {\bibinfo {volume} {85}},\
  \bibinfo {pages} {041122} (\bibinfo {year} {2012}{\natexlab{a}})}\BibitemShut
  {NoStop}%
\bibitem [{\citenamefont {Arenas}\ and\ \citenamefont
  {Barci}(2012{\natexlab{b}})}]{Arenas2012-2}%
  \BibitemOpen
  \bibfield  {author} {\bibinfo {author} {\bibfnamefont {Z.~G.}\ \bibnamefont
  {Arenas}}\ and\ \bibinfo {author} {\bibfnamefont {D.~G.}\ \bibnamefont
  {Barci}},\ }\href {http://stacks.iop.org/1742-5468/2012/i=12/a=P12005}
  {\bibfield  {journal} {\bibinfo  {journal} {Journal of Statistical Mechanics:
  Theory and Experiment}\ }\textbf {\bibinfo {volume} {2012}},\ \bibinfo
  {pages} {P12005} (\bibinfo {year} {2012}{\natexlab{b}})}\BibitemShut
  {NoStop}%
\bibitem [{\citenamefont {Moreno}\ \emph {et~al.}(2015)\citenamefont {Moreno},
  \citenamefont {Arenas},\ and\ \citenamefont {Barci}}]{Miguel2015}%
  \BibitemOpen
  \bibfield  {author} {\bibinfo {author} {\bibfnamefont {M.~V.}\ \bibnamefont
  {Moreno}}, \bibinfo {author} {\bibfnamefont {Z.~G.}\ \bibnamefont {Arenas}},
  \ and\ \bibinfo {author} {\bibfnamefont {D.~G.}\ \bibnamefont {Barci}},\
  }\href {\doibase 10.1103/PhysRevE.91.042103} {\bibfield  {journal} {\bibinfo
  {journal} {Phys. Rev. E}\ }\textbf {\bibinfo {volume} {91}},\ \bibinfo
  {pages} {042103} (\bibinfo {year} {2015})}\BibitemShut {NoStop}%
\bibitem [{\citenamefont {Aron}\ \emph {et~al.}(2016)\citenamefont {Aron},
  \citenamefont {Barci}, \citenamefont {Cugliandolo}, \citenamefont {Arenas},\
  and\ \citenamefont {Lozano}}]{ArBaCuZoGus2016}%
  \BibitemOpen
  \bibfield  {author} {\bibinfo {author} {\bibfnamefont {C.}~\bibnamefont
  {Aron}}, \bibinfo {author} {\bibfnamefont {D.~G.}\ \bibnamefont {Barci}},
  \bibinfo {author} {\bibfnamefont {L.~F.}\ \bibnamefont {Cugliandolo}},
  \bibinfo {author} {\bibfnamefont {Z.~G.}\ \bibnamefont {Arenas}}, \ and\
  \bibinfo {author} {\bibfnamefont {G.~S.}\ \bibnamefont {Lozano}},\ }\href
  {http://stacks.iop.org/1742-5468/2016/i=5/a=053207} {\bibfield  {journal}
  {\bibinfo  {journal} {Journal of Statistical Mechanics: Theory and
  Experiment}\ }\textbf {\bibinfo {volume} {2016}},\ \bibinfo {pages} {053207}
  (\bibinfo {year} {2016})}\BibitemShut {NoStop}%
\bibitem [{\citenamefont {Zinn-Justin}(2002)}]{Zinn-Justin}%
  \BibitemOpen
  \bibfield  {author} {\bibinfo {author} {\bibfnamefont {J.}~\bibnamefont
  {Zinn-Justin}},\ }\href@noop {} {\emph {\bibinfo {title} {Quantum field
  theory and critical phenomena}}}\ (\bibinfo  {publisher} {Oxford University
  Press},\ \bibinfo {address} {USA},\ \bibinfo {year} {2002})\BibitemShut
  {NoStop}%
\bibitem [{\citenamefont {Caroli}\ \emph {et~al.}(1979)\citenamefont {Caroli},
  \citenamefont {Caroli},\ and\ \citenamefont {Roulet}}]{Caroli1979}%
  \BibitemOpen
  \bibfield  {author} {\bibinfo {author} {\bibfnamefont {B.}~\bibnamefont
  {Caroli}}, \bibinfo {author} {\bibfnamefont {C.}~\bibnamefont {Caroli}}, \
  and\ \bibinfo {author} {\bibfnamefont {B.}~\bibnamefont {Roulet}},\ }\href
  {\doibase 10.1007/BF01009609} {\bibfield  {journal} {\bibinfo  {journal}
  {Journal of Statistical Physics}\ }\textbf {\bibinfo {volume} {21}},\
  \bibinfo {pages} {415} (\bibinfo {year} {1979})}\BibitemShut {NoStop}%
\bibitem [{\citenamefont {Caroli}\ \emph {et~al.}(1981)\citenamefont {Caroli},
  \citenamefont {Caroli},\ and\ \citenamefont {Roulet}}]{Caroli1981}%
  \BibitemOpen
  \bibfield  {author} {\bibinfo {author} {\bibfnamefont {B.}~\bibnamefont
  {Caroli}}, \bibinfo {author} {\bibfnamefont {C.}~\bibnamefont {Caroli}}, \
  and\ \bibinfo {author} {\bibfnamefont {B.}~\bibnamefont {Roulet}},\ }\href
  {\doibase 10.1007/BF01106788} {\bibfield  {journal} {\bibinfo  {journal}
  {Journal of Statistical Physics}\ }\textbf {\bibinfo {volume} {26}},\
  \bibinfo {pages} {83} (\bibinfo {year} {1981})}\BibitemShut {NoStop}%
\bibitem [{\citenamefont {S\'a~Borges}\ \emph {et~al.}(1988)\citenamefont
  {S\'a~Borges}, \citenamefont {Epele}, \citenamefont {Fanchiotti},
  \citenamefont {Garc\'{\i}a~Canal},\ and\ \citenamefont {Simo}}]{Epele1988}%
  \BibitemOpen
  \bibfield  {author} {\bibinfo {author} {\bibfnamefont {J.}~\bibnamefont
  {S\'a~Borges}}, \bibinfo {author} {\bibfnamefont {L.~N.}\ \bibnamefont
  {Epele}}, \bibinfo {author} {\bibfnamefont {H.}~\bibnamefont {Fanchiotti}},
  \bibinfo {author} {\bibfnamefont {C.~A.}\ \bibnamefont {Garc\'{\i}a~Canal}},
  \ and\ \bibinfo {author} {\bibfnamefont {F.~R.~A.}\ \bibnamefont {Simo}},\
  }\href {\doibase 10.1103/PhysRevA.38.3101} {\bibfield  {journal} {\bibinfo
  {journal} {Phys. Rev. A}\ }\textbf {\bibinfo {volume} {38}},\ \bibinfo
  {pages} {3101} (\bibinfo {year} {1988})}\BibitemShut {NoStop}%
\bibitem [{\citenamefont {It\^{o}}(1951)}]{Ito1951}%
  \BibitemOpen
  \bibfield  {author} {\bibinfo {author} {\bibfnamefont {K.}~\bibnamefont
  {It\^{o}}},\ }\href@noop {} {\bibfield  {journal} {\bibinfo  {journal} {J.
  Math. Soc. Japan}\ }\textbf {\bibinfo {volume} {3}},\ \bibinfo {pages} {157}
  (\bibinfo {year} {1951})}\BibitemShut {NoStop}%
\bibitem [{\citenamefont {H{\"a}nggi}(1978)}]{Hanggi1978}%
  \BibitemOpen
  \bibfield  {author} {\bibinfo {author} {\bibfnamefont {P.}~\bibnamefont
  {H{\"a}nggi}},\ }\href {\doibase 10.5169/seals-114941} {\bibfield  {journal}
  {\bibinfo  {journal} {Helv.\ Phys.\ Acta}\ }\textbf {\bibinfo {volume}
  {51}},\ \bibinfo {pages} {183} (\bibinfo {year} {1978})}\BibitemShut
  {NoStop}%
\bibitem [{\citenamefont {H{\"a}nggi}\ and\ \citenamefont
  {Thomas}(1982)}]{Hanggi1982}%
  \BibitemOpen
  \bibfield  {author} {\bibinfo {author} {\bibfnamefont {P.}~\bibnamefont
  {H{\"a}nggi}}\ and\ \bibinfo {author} {\bibfnamefont {H.}~\bibnamefont
  {Thomas}},\ }\href@noop {} {\bibfield  {journal} {\bibinfo  {journal} {Phys.
  Rep.}\ }\textbf {\bibinfo {volume} {88}},\ \bibinfo {pages} {207} (\bibinfo
  {year} {1982})}\BibitemShut {NoStop}%
\bibitem [{\citenamefont {Klimontovich}(1994)}]{Klimontovich}%
  \BibitemOpen
  \bibfield  {author} {\bibinfo {author} {\bibfnamefont {Y.~L.}\ \bibnamefont
  {Klimontovich}},\ }\href {\doibase 10.1070/PU1994v037n08ABEH000038}
  {\bibfield  {journal} {\bibinfo  {journal} {Physics-Uspekhi}\ }\textbf
  {\bibinfo {volume} {37}},\ \bibinfo {pages} {737} (\bibinfo {year}
  {1994})}\BibitemShut {NoStop}%
\bibitem [{\citenamefont {Onsager}\ and\ \citenamefont
  {Machlup}(1953)}]{Onsager1953}%
  \BibitemOpen
  \bibfield  {author} {\bibinfo {author} {\bibfnamefont {L.}~\bibnamefont
  {Onsager}}\ and\ \bibinfo {author} {\bibfnamefont {S.}~\bibnamefont
  {Machlup}},\ }\href {\doibase 10.1103/PhysRev.91.1505} {\bibfield  {journal}
  {\bibinfo  {journal} {Phys. Rev.}\ }\textbf {\bibinfo {volume} {91}},\
  \bibinfo {pages} {1505} (\bibinfo {year} {1953})}\BibitemShut {NoStop}%
\bibitem [{\citenamefont {Langouche}\ \emph {et~al.}(1979)\citenamefont
  {Langouche}, \citenamefont {Roekaerts},\ and\ \citenamefont
  {Tirapegui}}]{Langouche1979}%
  \BibitemOpen
  \bibfield  {author} {\bibinfo {author} {\bibfnamefont {F.}~\bibnamefont
  {Langouche}}, \bibinfo {author} {\bibfnamefont {D.}~\bibnamefont
  {Roekaerts}}, \ and\ \bibinfo {author} {\bibfnamefont {E.}~\bibnamefont
  {Tirapegui}},\ }\href {\doibase
  http://dx.doi.org/10.1016/0378-4371(79)90054-2} {\bibfield  {journal}
  {\bibinfo  {journal} {Physica A: Statistical Mechanics and its Applications}\
  }\textbf {\bibinfo {volume} {95}},\ \bibinfo {pages} {252 } (\bibinfo {year}
  {1979})}\BibitemShut {NoStop}%
\bibitem [{\citenamefont {Gitterman}(2005)}]{Gitterman}%
  \BibitemOpen
  \bibfield  {author} {\bibinfo {author} {\bibfnamefont {M.}~\bibnamefont
  {Gitterman}},\ }\href@noop {} {\emph {\bibinfo {title} {{The noisy
  oscillator: The first hundred years, from Einstein until now}}}}\ (\bibinfo
  {publisher} {World Scientific Publishing Co.},\ \bibinfo {year}
  {2005})\BibitemShut {NoStop}%
\bibitem [{\citenamefont {Cugliandolo}\ and\ \citenamefont
  {Lecomte}(2017)}]{Lecomte-Leticia2017}%
  \BibitemOpen
  \bibfield  {author} {\bibinfo {author} {\bibfnamefont {L.~F.}\ \bibnamefont
  {Cugliandolo}}\ and\ \bibinfo {author} {\bibfnamefont {V.}~\bibnamefont
  {Lecomte}},\ }\href {http://stacks.iop.org/1751-8121/50/i=34/a=345001}
  {\bibfield  {journal} {\bibinfo  {journal} {Journal of Physics A:
  Mathematical and Theoretical}\ }\textbf {\bibinfo {volume} {50}},\ \bibinfo
  {pages} {345001} (\bibinfo {year} {2017})}\BibitemShut {NoStop}%
\bibitem [{\citenamefont {{Moriconi}}(2004)}]{Luca2004}%
  \BibitemOpen
  \bibfield  {author} {\bibinfo {author} {\bibfnamefont {L.}~\bibnamefont
  {{Moriconi}}},\ }\href {\doibase 10.1119/1.1715108} {\bibfield  {journal}
  {\bibinfo  {journal} {American Journal of Physics}\ }\textbf {\bibinfo
  {volume} {72}},\ \bibinfo {pages} {1258} (\bibinfo {year} {2004})},\ \Eprint
  {http://arxiv.org/abs/physics/0402069} {physics/0402069} \BibitemShut
  {NoStop}%
\bibitem [{\citenamefont {Rubin}\ \emph {et~al.}(2014)\citenamefont {Rubin},
  \citenamefont {Pruessner},\ and\ \citenamefont {Pavliotis}}]{Rubin2014}%
  \BibitemOpen
  \bibfield  {author} {\bibinfo {author} {\bibfnamefont {K.~J.}\ \bibnamefont
  {Rubin}}, \bibinfo {author} {\bibfnamefont {G.}~\bibnamefont {Pruessner}}, \
  and\ \bibinfo {author} {\bibfnamefont {G.~A.}\ \bibnamefont {Pavliotis}},\
  }\href {http://stacks.iop.org/1751-8121/47/i=19/a=195001} {\bibfield
  {journal} {\bibinfo  {journal} {Journal of Physics A: Mathematical and
  Theoretical}\ }\textbf {\bibinfo {volume} {47}},\ \bibinfo {pages} {195001}
  (\bibinfo {year} {2014})}\BibitemShut {NoStop}%
\bibitem [{\citenamefont {Dunne}(2008)}]{Dunne2008}%
  \BibitemOpen
  \bibfield  {author} {\bibinfo {author} {\bibfnamefont {G.~V.}\ \bibnamefont
  {Dunne}},\ }\href {http://stacks.iop.org/1751-8121/41/i=30/a=304006}
  {\bibfield  {journal} {\bibinfo  {journal} {Journal of Physics A:
  Mathematical and Theoretical}\ }\textbf {\bibinfo {volume} {41}},\ \bibinfo
  {pages} {304006} (\bibinfo {year} {2008})}\BibitemShut {NoStop}%
\bibitem [{\citenamefont {Abramowitz}\ and\ \citenamefont
  {Stegun}(1964)}]{abramowitz-1964}%
  \BibitemOpen
  \bibfield  {author} {\bibinfo {author} {\bibfnamefont {M.}~\bibnamefont
  {Abramowitz}}\ and\ \bibinfo {author} {\bibfnamefont {I.~A.}\ \bibnamefont
  {Stegun}},\ }\href@noop {} {\emph {\bibinfo {title} {Handbook of Mathematical
  Functions with Formulas, Graphs, and Mathematical Tables}}},\ \bibinfo
  {edition} {{ninth Dover printing, tenth GPO printing}}\ ed.\ (\bibinfo
  {publisher} {Dover},\ \bibinfo {address} {New York},\ \bibinfo {year}
  {1964})\BibitemShut {NoStop}%
\bibitem [{\citenamefont {Gradshteyn}\ and\ \citenamefont
  {Ryzhik}(2007)}]{gradshteyn2007}%
  \BibitemOpen
  \bibfield  {author} {\bibinfo {author} {\bibfnamefont {I.~S.}\ \bibnamefont
  {Gradshteyn}}\ and\ \bibinfo {author} {\bibfnamefont {I.~M.}\ \bibnamefont
  {Ryzhik}},\ }\href@noop {} {\emph {\bibinfo {title} {Table of integrals,
  series, and products}}},\ \bibinfo {edition} {seventh}\ ed.\ (\bibinfo
  {publisher} {Elsevier/Academic Press, Amsterdam},\ \bibinfo {year} {2007})\
  pp.\ \bibinfo {pages} {xlviii+1171},\ \bibinfo {note} {translated from the
  Russian, Translation edited and with a preface by Alan Jeffrey and Daniel
  Zwillinger, With one CD-ROM (Windows, Macintosh and UNIX)}\BibitemShut
  {NoStop}%
\bibitem [{\citenamefont {Elizalde}\ \emph {et~al.}(1994)\citenamefont
  {Elizalde}, \citenamefont {Odintsov}, \citenamefont {Romeo}, \citenamefont
  {Bytsenko},\ and\ \citenamefont {Zerbini}}]{zfunction1994}%
  \BibitemOpen
  \bibfield  {author} {\bibinfo {author} {\bibfnamefont {E.}~\bibnamefont
  {Elizalde}}, \bibinfo {author} {\bibfnamefont {S.~D.}\ \bibnamefont
  {Odintsov}}, \bibinfo {author} {\bibfnamefont {A.}~\bibnamefont {Romeo}},
  \bibinfo {author} {\bibfnamefont {A.~A.}\ \bibnamefont {Bytsenko}}, \ and\
  \bibinfo {author} {\bibfnamefont {S.}~\bibnamefont {Zerbini}},\ }\href
  {\doibase 10.1142/2065} {\emph {\bibinfo {title} {Zeta Regularization
  Techniques with Applications}}}\ (\bibinfo  {publisher} {WORLD SCIENTIFIC},\
  \bibinfo {year} {1994})\ \Eprint
  {http://arxiv.org/abs/https://www.worldscientific.com/doi/pdf/10.1142/2065}
  {https://www.worldscientific.com/doi/pdf/10.1142/2065} \BibitemShut {NoStop}%
\end{thebibliography}

%

\end{document}